\documentclass[aps,prd,
groupaddress,twocolumn,nofootinbib]{revtex4-1}

\usepackage{amsmath}
\usepackage{amssymb}
\usepackage{graphicx}
\usepackage{mathrsfs}
\usepackage{hyperref}
\usepackage{braket}

\def\D{\mathrm{d}}
\def\e{\mathrm{e}}
\def\E{\mathrm{E}}
\def\I{\mathrm{i}}
\def\Ob{\mathcal{O}}
\def\NN{\mathcal{N}}

\usepackage{xcolor}

\begin{document}

\title{Relational observables, reference frames, and conditional probabilities}
 
 \author{Leonardo Chataignier}
 \email{lcmr@thp.uni-koeln.de}
 \affiliation{Institut f\"ur Theoretische Physik, Universit\"{a}t zu
   K\"{o}ln, Z\"{u}lpicher Stra\ss e 77, 50937 K\"{o}ln, Germany} 
 
\date{\today}

\begin{abstract}
We discuss the construction of relational observables in time-reparametrization invariant quantum mechanics and we argue that their physical interpretation can be understood in terms of conditional probabilities, which are defined from the solutions of the quantum constraint equation in a generalization of the Page-Wootters formalism. In this regard, we show how conditional expectation values of worldline tensor fields are related to quantum averages of suitably defined relational observables. We also comment on how the dynamics of these observables can be related to a notion of quantum reference frames. After presenting the general formalism, we analyze a recollapsing cosmological model, for which we construct unitarily evolving quantum relational observables. We conclude with some remarks about the relevance of these results for the construction and interpretation of diffeomorphism-invariant operators in quantum gravity.
\end{abstract}

\maketitle

\section{Introduction}
At present, there are many approaches to quantum gravity (see, for instance,~\cite{Oriti:book} and references therein). Among the various conceptual and technical difficulties that each of them faces, there are two pressing issues. The first is a matter of quantum foundations, which is the correct physical interpretation of the quantum states of geometry and matter: how are they related to probability distributions of histories or configurations of a gravitational system? The second is the precise definition of the quantum observables and their classical limit: assuming there is a physical Hilbert space of states, what operators act on this space and have a reasonable interpretation?

Over the years, several proposals have been made to (partially) resolve these issues. For example, it has been argued that quantum states in generally covariant systems can be used to define conditional probabilities~\cite{PW1,PW2,PW3,PW4,PW5,PW6,PW7,Dolby:2004,Hoehn:Trinity}, or that the probabilistic interpretation of wave functions is justified within (a generalization of) the consistent or decoherent histories formalism (see~\cite{DH2,DH4,DH5,DH6,DH8} and references therein). There are also other related approaches, such as Rovelli's relational quantum mechanics~\cite{relQM}. Alternatively, one may adopt the de Broglie-Bohm theory (see~\cite{dBB1,dBB2} and references therein) and consider that solutions to the quantum constraint equation give rise to trajectories in configuration space, whereas probabilities would play a secondary role.

Moreover, the construction of relevant observables in quantum gravity and cosmology has been a topic of active research. Rovelli has argued that the relevant observables are ``evolving constants of motion''~\cite{Rovelli:1990-1,Rovelli:1990-2,Rovelli:1991}, which are diffeomorphism-invariant extensions of tensor fields~\cite{Woodard:1985,Teitelboim:1992,HT:book,Woodard:1993,Dittrich:2004,Dittrich:2005,Chataig:2019} that capture the relational evolution between different quantities. For this reason, they can also be referred to as ``relational observables''~\cite{Tambornino:2012}. The correct way to construct these quantities in the quantum theory has been a subject of debate~\cite{Dittrich:2004,Dittrich:2005,Tambornino:2012,Hoehn:2018-1,Hoehn:2018-2,Hoehn:2019}. Recently, the author has proposed a systematic method to construct quantum relational observables in generally covariant quantum mechanics~\cite{Chataig:2019} (see also~\cite{Hoehn:Trinity} for another recent, similar approach).

In the present article, we relate the two issues (the probabilistic interpretation of the physical states and the construction of relevant observables in quantum gravity) by arguing that relational observables are conditional quantities and that their quantum dynamics can be encoded in conditional probability amplitudes. We first make this argument in general, i.e., for any time-reparametrization invariant mechanical theory. We then support this claim by analyzing the construction of quantum relational observables in a cosmological model. For simplicity, we tackle these problems in the framework of canonical quantum gravity using metric variables. This might not be the most fundamental approach to quantum gravity, but it is sufficient for our purposes since it has a straightforward classical limit~\cite{Kiefer:book}. We use a generalization of the formalism presented by the author in~\cite{Chataig:2019} and the reader is referred to~\cite{Hoehn:Trinity} for a discussion of conditional probabilities in the subsequent, related approach that was developed there. The restriction to mechanical (symmetry-reduced) models is made to avoid the field-theoretic issues of regularization and possible presence of anomalies in the quantum theory. In our conclusions, we discuss the relevance of our results to the construction of relational observables in the field-theoretic case.

The article is organized into four sections. In Sec.~\ref{sec:general}, we review and generalize the method of construction of quantum relational observables which was presented by the author in~\cite{Chataig:2019} and we relate it to conditional probabilities, which are defined from the solutions of the quantum constraint equation. We also comment on the relation of our approach with previous proposals, specifically the Page-Wootters formalism~\cite{PW1,PW2,PW3,PW4,PW5,PW6,PW7,Dolby:2004,Hoehn:Trinity} and the relativization maps and $G$-twirl operations used in the study of quantum reference frames in the context of quantum foundations and quantum information science (see~\cite{Hoehn:Trinity,QRef1,QRef3,QRef4,QRef5} and references therein). We then analyze the concrete example of a Friedmann–Lema\^{i}tre–Robertson–Walker (FLRW) model in Sec.~\ref{sec:flrw}. We construct the relevant relational observables, which have the expected unitary evolution governed by a physical Hamiltonian operator. Finally, in Sec.~\ref{sec:conclusions}, we restate our results and present our conclusions.

\section{\label{sec:general}The General Framework}
In this section, we generalize the formalism presented in~\cite{Chataig:2019} for the construction of quantum relational observables in time-reparametrization invariant mechanical systems. The reader is directed to~\cite{Chataig:2019} for further details and references to the earlier literature. Furthermore, we argue that the physical interpretation of relational observables should be tied to the use of conditional probabilities in the quantum theory and we relate the matrix elements of quantum relational observables to conditional expectation values. A concrete example of the general formalism presented in this section will be analyzed in Sec.~\ref{sec:flrw}.

\subsection{Classical theory}\label{sec:classical-general}
We consider a general mechanical system comprised of the degrees of freedom $p_i(\tau), q^i(\tau)$, $(i = 1,\ldots,n)$ and the action in Hamiltonian form
\begin{equation}\label{eq:action-general}
    S = \int_a^b\D\tau\ \left[p_i(\tau)\dot{q}^i(\tau)-e(\tau)C(q(\tau),p(\tau))\right] \ ,
\end{equation}
where $\cdot\equiv\frac{\D}{\D\tau}$ and a summation over the index $i$ is implied. We assume that $p_i(\tau), q^i(\tau)$ transform as scalars under $\tau$-reparametrizations. The field $e(\tau)$ is an arbitrary multiplier, which transforms as a scalar density under $\tau$-reparametrizations,
\begin{equation}\label{eq:general-einbein}
    e(\tau) = e'(\tau')\frac{\D\tau'}{\D\tau} \ .
\end{equation}
For this reason, $e(\tau)$ can be interpreted as the einbein on the $\tau$-manifold (the ``worldline''). Its field equation is a constraint
\begin{equation}\label{eq:classical-C-general}
    C(q(\tau),p(\tau)) = 0 \ .
\end{equation}
From~(\ref{eq:action-general}), we see that the canonical Hamiltonian is $e(\tau)C(q(\tau),p(\tau))$, and it vanishes on shell, i.e., on the constraint surface defined by~(\ref{eq:classical-C-general}). Under an infinitesimal reparametrization, $\tau\mapsto\tau-\epsilon(\tau)$, the action~(\ref{eq:action-general}) remains invariant if $\epsilon(a) = \epsilon(b) = 0$. In general ($\epsilon(a),\epsilon(b)\neq0$), the action can be made invariant with the addition of certain boundary terms~\cite{HT:Vergara}. On the other hand, a general phase-space function $f(q(\tau),p(\tau))$ is not invariant, as it transforms as
\begin{equation}
    \delta_{\epsilon(\tau)}f = \epsilon(\tau)\frac{\D f}{\D\tau} = \epsilon(\tau)\left\{f,e(\tau)C\right\} \ ,
\end{equation}
where $\{\cdot,\cdot\}$ is the Poisson bracket. For reasons which will become clear in what follows, we will be interested in invariant operators (also called Dirac observables) in the quantum theory. Therefore, it is important to understand what kind of invariant objects one can construct already at the classical level.

The object~\cite{Chataig:2019,Marolf:1995}
\begin{equation}\label{eq:general-invariant}
    \Ob_{\omega} :=\int_{-\infty}^{\infty}\D\tau\ \omega(\tau) \ ,
\end{equation}
where $\omega(\tau)$ is a worldline one-form, is a well-defined invariant, provided the integral converges. Indeed, Eq.~(\ref{eq:general-invariant}) is independent of the choice of parametrization $\tau$. Can we find objects of form of~(\ref{eq:general-invariant}) that have a clear physical interpretation? The answer is yes. To do so, we follow~\cite{Chataig:2019} and define a new coordinate $s$ on the worldline via the ``gauge condition''
\begin{equation}\label{eq:gauge-condition-general}
    \chi(q(\tau),p(\tau)) = s \ ,
\end{equation}
where $\chi$ is a worldline scalar that satisfies
\begin{equation}\label{eq:admissible-general}
   \frac{\D\chi}{\D\tau} = \{\chi,e(\tau)C\}\neq 0
\end{equation}
in certain regions of phase space. Due to~(\ref{eq:admissible-general}), we can locally solve~(\ref{eq:gauge-condition-general}) for $\tau$ to obtain
\begin{equation}\label{eq:phi-diffeo}
    \tau = \phi(q(0),p(0),s) \equiv \phi(s) \ ,
\end{equation}
where $\phi$ defines a worldline diffeomorphism. Given a worldline scalar $f(\tau)$, we can consider its pullback
\begin{equation}\label{eq:general-pullback}
    \Ob[f|\chi = s]:=\phi^*f = \left.f(\tau)\right|_{\tau = \phi} \ .
\end{equation}
It is straightforward to verify that~(\ref{eq:general-pullback}) can also be written as~\cite{Chataig:2019}
\begin{equation}\label{eq:relObs-general}
    \Ob[f|\chi = s] = \Delta_{\chi}\int_{-\infty}^{\infty}\D\tau\ \delta(\chi(q(\tau),p(\tau))-s)f(\tau) \ ,
\end{equation}
where\footnote{A word of caution: the notation used here is different from the one used in~\cite{Chataig:2019}, where $\Delta_{\chi}$ was used to denote $\frac{\D\chi}{\D\tau}$, and $\left|\Delta_{\chi}^{\Ob}\right|^{-1}$ was used to denote the quantity that appears in the right-hand side of Eq.~(\ref{eq:FP-general}) in the present article. Here, we use a simpler notation for convenience.}
\begin{equation}\label{eq:FP-general}
    \Delta_{\chi}^{-1}:=\int_{-\infty}^{\infty}\D\tau\ \delta(\chi(q(\tau),p(\tau))-s) \ .
\end{equation}
Equation~(\ref{eq:relObs-general}) is of the same form as~(\ref{eq:general-invariant}) and it defines a relational observable, which is a diffeomorphism-invariant quantity that represents the value of $f(\tau)$ when $\chi(q(\tau),p(\tau)) = s$. In other words, the physical interpretation of the observable given in~(\ref{eq:relObs-general}) is that it yields the value of $f(\tau)$ in relation to a given value of $\chi(q(\tau),p(\tau))$, the level sets of which are used to define a new coordinate $s$. Evidently, $f(\tau)$ can be replaced by a general phase-space function $f(q(\tau),p(\tau))$.

It should be noted that, in the most general case, it may be necessary to modify both~(\ref{eq:relObs-general}) and~(\ref{eq:FP-general}) with the substitution
\begin{equation}\label{eq:non-monotonic-subs}
\delta(\chi(q(\tau),p(\tau))-s) \mapsto \delta(\chi(q(\tau),p(\tau))-s)g(q(\tau),p(\tau)) \ ,
\end{equation}
where $g(q(\tau),p(\tau))$ is a phase-space function that vanishes outside of a given phase-space region where~(\ref{eq:admissible-general}) is fulfilled. Indeed, a general choice of $\chi(q(\tau),p(\tau))$ is not a monotonic function of $\tau$ and will feature turning points where~(\ref{eq:admissible-general}) does not hold. The role of $g(q(\tau),p(\tau))$ is thus to restrict a general $\chi(q(\tau),p(\tau))$ to a region where it is admissible, i.e., where~(\ref{eq:admissible-general}) is satisfied. Only then are the field-dependent diffeomorphism~(\ref{eq:phi-diffeo}) and, consequently, the observables~(\ref{eq:relObs-general}) well-defined. For example, in certain cases, one could set $g(q(\tau),p(\tau)) = \theta\left(\pm\frac{\D\chi}{\D\tau}\right)$ to restrict the gauge condition to regions where its derivative is either positive or negative ($\theta(x)$ is the Heaviside step function). A similar observation was made in~\cite{Chataig:2019}. For simplicity, we will omit the use of $g(q(\tau),p(\tau))$ in what follows. We briefly comment on the effect of its inclusion in the formalism in Sec.~\ref{sec:relObs-I} (see footnote~\ref{foot:omega}).

In the particular case in which $f(\tau) = 1$, we obtain
\begin{equation}\label{eq:FP-general-0}
    \Ob[1|\chi = s] = 1 \ ,
\end{equation}
due to~(\ref{eq:FP-general}). This is the usual Faddeev-Popov resolution of the identity~\cite{FP-1,FP-2}, more commonly expressed by rewriting~(\ref{eq:FP-general}) as
\begin{equation}\label{eq:FP-general-1}
    1 = \Delta_{\chi}\int_{-\infty}^{\infty}\D\tau\ \delta(\chi(q(\tau),p(\tau))-s) \ .
\end{equation}
Moreover, if $f(q(\tau),p(\tau)) = \chi(q,(\tau)p(\tau))$, we find from~(\ref{eq:relObs-general}) that $\Ob[\chi|\chi = s] = s$. It is also possible to consider relational observables associated with worldline one-forms, but we do not consider this here (see~\cite{Chataig:2019}). This construction of relational observables was extensively analyzed in~\cite{Chataig:2019}, where it was shown that the observables satisfy gauge-fixed equations of motion with respect to the new coordinate $s$. In Sec.~\ref{sec:flrw}, we present a concrete example of such observables.

The on-shell invariant observables constructed from general phase-space functions $f(q(\tau),p(\tau))$ can be seen as functions on the so-called physical or reduced phase space~\cite{HT:book}, which is the space of orbits generated by the constraint~(\ref{eq:classical-C-general}) (or, more precisely, by the related gauge generator~\cite{Chataig:2019}) and it can be labeled by a complete set of independent on-shell invariants of the form $\Ob[f|\chi = s_0] = \phi^*f$ [cf.~(\ref{eq:general-pullback})] for a fixed value of $s_0$. As explained in~\cite{Chataig:2019}, these quantities can be seen as diffeomorphism-invariant extensions of the initial values of the worldline scalars.

More precisely, we note that choosing an admissible gauge condition $\chi(q(\tau),p(\tau))$ [cf.~(\ref{eq:admissible-general})] corresponds to selecting a particular ``time reference frame'' for the reparametrization-invariant system, i.e., the frame defined by the level sets of the scalar $\chi(q(\tau),p(\tau))$, and that $\Ob[f|\chi=s]$ corresponds to (a diffeomorphism-invariant extension of) the worldline scalar $f$ described in that frame. The evolution of worldline scalars $f(q(\tau),p(\tau))$ in a given frame is governed by a quantity $H_{\chi}^{\sigma}$, called the physical or reduced phase-space Hamiltonian, which corresponds to (an invariant extension of) the opposite of $p_{\chi}$, the canonical momentum conjugate to $\chi$. This can be justified from the on-shell action in the following way. We assume that $\chi(q(\tau),p(\tau))$ is one of the configuration variables (after a canonical transformation, if necessary) and that we can solve the constraint equation~(\ref{eq:classical-C-general}) for $p_{\chi}$ to find
\begin{equation}\label{eq:classical-freq-sectors}
p_{\chi} = -H_{\chi}^{\sigma}(q(\tau),p(\tau))\equiv-H_{\chi}^{\sigma}(\tau) \ ,
\end{equation}
where $\sigma$ is a possible multiplicity of the solution. This multiplicity depends on the form of the original constraint $C$ (e.g., the positive and negative frequency sectors which result from the quadratic constraint for the free relativistic particle~\cite{Chataig:2019}). From~(\ref{eq:action-general}) and~(\ref{eq:classical-freq-sectors}), we obtain the on-shell action
\begin{equation}\label{eq:on-shell-action-general}
S_{C = 0} = \int_a^b\D\tau \left(\sum_{i\neq\chi}p_i(\tau)\dot{q}^i(\tau)-H_{\chi}^{\sigma}(\tau)\dot{\chi}(\tau)\right) \ ,
\end{equation}
where we have neglected a possible boundary term needed to make the action reparametrization-invariant~\cite{HT:Vergara}, since it will not concern us here. If we now reparametrize the integrand of~(\ref{eq:on-shell-action-general}) using~(\ref{eq:phi-diffeo}), we obtain
\begin{equation}\label{eq:on-shell-action-general-2}
S_{C = 0} = \int_{\phi^{-1}(a)}^{\phi^{-1}(b)}\!\!\!\D s \left[\sum_{i\neq\chi}p_i(\phi(s))\frac{\D q^i(\phi(s))}{\D s}-H_{\chi}^{\sigma}(\phi(s))\right] ,
\end{equation}
which is the action of an ordinary (unconstrained) physical system that describes the evolution of the invariants $\Ob[q|\chi = s] = \phi^*q$ with respect to the gauge-fixed time parameter $s$. The physical Hamiltonian is $H_{\chi}^{\sigma}(\phi(s)) = \phi^*H_{\chi}^{\sigma}$, i.e., it can also in principle be described as a diffeomorphism-invariant quantity (for each fixed value of $s$).

In Sec.~\ref{sec:QT}, we will discuss possible definitions of quantum relational observables and argue that they are related to a notion of ``quantum reference frames'' that can, furthermore, be physically interpreted with the use of conditional probabilities.

\subsection{\label{sec:QT}Quantum theory}
\subsubsection{\label{sec:QT:Hilbert}The physical Hilbert space}
To construct the quantum theory, we consider an auxiliary (off-shell) Hilbert space in which $\hat{p}_i,\hat{q}^i$ are self-adjoint operators with respect to an auxiliary inner product $\braket{\cdot|\cdot}$. We then promote the classical constraint~(\ref{eq:classical-C-general}) to a linear operator $\hat{C}$ which is self-adjoint with respect to $\braket{\cdot|\cdot}$. Its eigenstates $\ket{E,\bf k}$ satisfy
\begin{align*}
    \hat{C}\ket{E,\bf k} &= E\ket{E,\bf k} \ ,\\
    \braket{E',{\bf k'}|E,\bf k} &= \delta(E',E)(E,{\bf k'}|E,\bf k) \ ,
\end{align*}
where we defined the induced inner product~\cite{Rieffel:1974,HT-SUSY:1982,Landsman:1995,Marolf:1995-4,Marolf:1997,Marolf:2000,Chataig:2019}
\begin{equation}\label{eq:induced-IP}
    (E,{\bf k'}|E,\bf k) := \delta(k',k) \ .
\end{equation}
Here, ${\bf k}$ is a degeneracy that can be seen as a label in the reduced configuration space, and $\delta(\cdot,\cdot)$ is a Kronecker or Dirac delta if the spectrum of $\hat{C}$ is, respectively, discrete or continuous. The induced inner product~(\ref{eq:induced-IP}) is well-defined even for on-shell states (for which $E' = E = 0$). The physical (on-shell) Hilbert space is the vector space of superpositions of $\ket{E = 0,\bf k}$ that are square-integrable with respect to the induced inner product $(\cdot|\cdot)$. Given two on-shell states
\begin{equation}\label{eq:on-shell-state-general}
    \ket{\Psi^{(1,2)}} = \sum_{\bf k}\Psi^{(1,2)}({\bf k})\ket{E = 0,\bf k} \ ,
\end{equation}
we can compute their induced overlap
\begin{equation}\label{eq:induced-prod-general}
    (\Psi^{(1)}|\Psi^{(2)}) = \sum_{\bf k}\ \overline{\Psi}^{(1)}({\bf k})\Psi^{(2)}({\bf k}) \ .
\end{equation}
The summation over ${\bf k}$ should be replaced by an integral if ${\bf k}$ is a continuous index. In what follows, we assume that $E$ is a continuous label that takes values over $\mathbb{R}$. The improper projectors onto a given eigenspace of $\hat{C}$ are
\begin{equation}\label{eq:general-projectors}
    \hat{P}_E = \sum_{\bf k}\ \ket{E,\bf k}\bra{E,\bf k} \ ,
\end{equation}
and they satisfy
\begin{align}
    \hat{P}_{E'}\hat{P}_E &= \delta(E'-E)\hat{P}_E \ ,\\
    \hat{P}_{E}\ket{\Psi^{(1,2)}} &= \delta(E)\ket{\Psi^{(1,2)}} \ . \label{eq:pre-fis-ID}
\end{align}
If we denote by $\bullet$ the action of $\hat{P}_{E = 0}$ with respect to the induced inner product~(\ref{eq:induced-IP}), then~(\ref{eq:pre-fis-ID}) implies that
\begin{equation}\label{eq:fis-ID}
    \hat{P}_{E = 0}\bullet\ket{\Psi^{(1,2)}} = \ket{\Psi^{(1,2)}} \ ;
\end{equation}
i.e., $\hat{P}_{E = 0}$ acts as the identity in the physical Hilbert space equipped with the induced inner product.

The inner product~(\ref{eq:induced-prod-general}) leads to the ``group-averaging'' inner product~\cite{Marolf:1995-4,Marolf:2000} in the following way. We can write
\begin{equation}\label{eq:group-av-general}
\begin{aligned}
    \hat{P}_{E = 0} &= \int_{-\infty}^{\infty}\D E\ \delta(E)\sum_{\bf k}\ket{E,\bf k}\bra{E,\bf k}\\
    &=\frac{1}{2\pi\hbar}\int\D\tau\D E\ \e^{\frac\I\hbar\tau\hat{C}}\sum_{\bf k}\ket{E,\bf k}\bra{E,\bf k}\\
    &=\frac{1}{2\pi\hbar}\int_{-\infty}^{\infty}\D\tau\ \e^{\frac\I\hbar\tau\hat{C}} \ ,
\end{aligned}
\end{equation}
where we used the fact that the states $\ket{E,\bf k}$ form a complete orthonormal system. If we assume that there exist off-shell states $\ket{\psi^{(1,2)}}$ such that
\begin{equation}\notag
    \Psi^{(1,2)}({\bf k}) = \braket{E = 0,{\bf k}|\psi^{(1,2)}} \ ,
\end{equation}
then from~(\ref{eq:on-shell-state-general}) we obtain
\begin{equation}\label{eq:phys-project-aux}
    \ket{\Psi^{(1,2)}} = \hat{P}_{E = 0}\ket{\psi^{(1,2)}} \ ,
\end{equation}
and from~(\ref{eq:induced-prod-general}),
\begin{equation}\label{eq:P-prod-general}
    (\Psi^{(1)}|\Psi^{(2)}) = \left<\psi^{(1)}\left|\hat{P}_{E = 0}\right|\psi^{(2)}\right> \ .
\end{equation}
Using~(\ref{eq:group-av-general}) in~(\ref{eq:P-prod-general}), we find the usual expression for the ``group averaging'' inner product
\begin{equation}
    (\Psi^{(1)}|\Psi^{(2)}) = \left<\psi^{(1)}\left|\frac{1}{2\pi\hbar}\int_{-\infty}^{\infty}\D\tau\ \e^{\frac\I\hbar\tau\hat{C}}\right|\psi^{(2)}\right> \ .
\end{equation}

\subsubsection{On-shell operators}
As~(\ref{eq:general-invariant}) is independent of the choice of parametrization $\tau$, we can choose $e(\tau) = 1$~[cf.~(\ref{eq:general-einbein})] and promote~(\ref{eq:general-invariant}) to the operator
\begin{equation}\label{eq:general-inv-quantum}
    \hat{\Ob}_{\omega}:=\int_{-\infty}^{\infty}\D\tau\ \e^{\frac\I\hbar\tau\hat{C}}\ \hat{\omega}\ \e^{-\frac\I\hbar\tau\hat{C}} \ ,
\end{equation}
assuming $\hat{\omega}$ can be defined with an appropriate choice of factor ordering. Why are these invariant objects interesting in the quantum theory? The reason is that the relevant operators are those which correspond to linear transformations between on-shell states. Such operators act only on the physical Hilbert space (via the induced inner product~(\ref{eq:fis-ID})) and they have the general form
\begin{equation}\label{eq:on-shell-op}
    \hat{\Ob} = \sum_{\bf k', k}\ \Ob({\bf k'},{\bf k})\ket{E = 0,\bf k'}\bra{E = 0,\bf k} \ .
\end{equation}
Let us now see that~(\ref{eq:general-inv-quantum}) leads precisely to objects of this kind (see also~\cite{Marolf:1995,Chataig:2019}). We first note that the matrix element of~(\ref{eq:general-inv-quantum}) between two eigenstates of the constraint operator with respect to the auxiliary inner product is
\begin{equation}
    \left<E',{\bf k'}\left|\hat{\Ob}_{\omega}\right|E,{\bf k}\right> = \delta(E'-E)\left(E,{\bf k'}\left|\hat{\Ob}_{\omega}\right|E,{\bf k}\right) \ ,
\end{equation}
where
\begin{equation}\label{eq:induced-matrix-omega-general}
    \left(E,{\bf k'}\left|\hat{\Ob}_{\omega}\right|E,{\bf k}\right) = 2\pi\hbar\left<E,{\bf k'}\left|\hat{\omega}\right|E,{\bf k}\right>
\end{equation}
is the induced matrix element of~(\ref{eq:general-inv-quantum}). In the physical Hilbert space spanned by the eigenstates $\ket{E = 0,\bf k}$, we therefore can write the induced matrix elements of~(\ref{eq:general-inv-quantum}) as
\begin{equation}\label{eq:on-shell-kernel-inv}
    \Ob_{\omega}({\bf k'},{\bf k}) := 2\pi\hbar\left<E = 0,{\bf k'}\left|\hat{\omega}\right|E = 0,{\bf k}\right> \ .
\end{equation}
In this way, the invariant operator $\hat{\Ob}_{\omega}$ can be represented in the physical Hilbert space as the operator~(\ref{eq:on-shell-op}) with $\Ob({\bf k'},{\bf k})$ given by~(\ref{eq:on-shell-kernel-inv}). Using~(\ref{eq:general-projectors}),~(\ref{eq:on-shell-op}) and~(\ref{eq:on-shell-kernel-inv}), we can also represent $\hat{\Ob}_{\omega}$ as the on-shell operator
\begin{equation}\label{eq:general-on-shell-PoP}
    \hat{\Ob}_{\omega} = 2\pi\hbar\ \hat{P}_{E = 0}\ \hat{\omega}\ \hat{P}_{E = 0} \ .
\end{equation}
Thus, the invariant operators~(\ref{eq:general-inv-quantum}) can be represented as linear transformations in the physical Hilbert space. Nevertheless, not all invariant operators are of immediate physical significance. As we have argued in the classical theory, the relational observables are invariants which have a clear physical interpretation. Our task is then to find and interpret the equivalent of~(\ref{eq:relObs-general}) in the quantum theory.

\subsubsection{\label{sec:relObs-I}Quantum relational observables I}
We now examine a slightly refined version of the method presented in~\cite{Chataig:2019} for the construction of quantum relational observables. Afterwards, we present a different perspective on this formalism (as well as an alternative choice of operator ordering) in Sec.~\ref{sec:relObs-II}. In Secs.~\ref{sec:QRef} and~\ref{sec:CP-general}, respectively, we comment on the physical interpretation of the constructed observables and we show how their matrix elements can be related to conditional probabilities, with a view to obtaining a clear physical interpretation of the formalism.

To quantize~(\ref{eq:relObs-general}), we must determine a choice of operator ordering. In~\cite{Chataig:2019}, it was argued that the ordering should be chosen so as to yield an operator-version of the Faddeev-Popov resolution of the identity~(\ref{eq:FP-general-1}). In order to achieve this, we assume that we can promote the classical gauge condition $\chi$ to an operator $\hat{\chi}$ which is self-adjoint with respect to the auxiliary inner product and which has the eigenstates $\ket{\chi,\bf n}$, where ${\bf n}$ indicates possible (discrete or continuous) degeneracies. In general, the on-shell components of these eigenstates, $\hat{P}_{E = 0}\ket{\chi,\bf n}$, do not form a complete orthonormal system in the physical Hilbert space. To obtain such a system, we define a self-adjoint\footnote{\label{foot:omega}As we will see in the discussion following~(\ref{eq:relObs-quantum-3}), $\hat{\Omega}_{\chi}^{\sigma}$ can be interpreted as the on-shell quantum analogue of the square root of the Faddeev-Popov determinant $\Delta_{\chi}$. We have seen that in the most general classical case, it may be necessary to make the substitution~(\ref{eq:non-monotonic-subs}) to ensure that the gauge condition is admissible. In the quantum theory, it may be necessary to consider $\hat{\Omega}_{\chi}^{\sigma}$ that is not self-adjoint, in which case it would be a quantum analogue of the square root of the product $g(q(\tau),p(\tau))\Delta_{\chi}$ with a suitable choice of ordering. We do not consider this complication here.} on-shell operator
\begin{equation}\label{eq:omega}
\hat{\Omega}_{\chi}^{\sigma} = \sum_{\bf k', k} \Omega^{\sigma}_{\chi}({\bf k', k})\ket{E = 0, \bf k'}\bra{E = 0, \bf k}
\end{equation}
by requiring that it satisfies the orthonormality conditions
\begin{align}
&2\pi\hbar\sum_{\sigma}\hat{\Omega}_{\chi}^{\sigma}\hat{P}_{\chi = s}\hat{\Omega}_{\chi}^{\sigma} = \hat{P}_{E = 0} \ , \label{eq:conditions-omega-1}\\
&\left<\chi = s, {\bf n'}\left|\hat{\Omega}_{\chi}^{\sigma'}\bullet\hat{\Omega}_{\chi}^{\sigma}\right|\chi = s, {\bf n}\right> = \frac{\delta_{\sigma',\sigma}\delta({\bf n', n})}{2\pi\hbar}\ ,\label{eq:conditions-omega-2}
\end{align}
where
\begin{equation}\label{eq:Pchi-general}
\hat{P}_{\chi = s}:=\sum_{\bf n}\ket{\chi = s,\bf n}\bra{\chi = s,\bf n} \ , 
\end{equation}
and $\sigma$ is a possible discrete degeneracy related to the classical multiplicity in~(\ref{eq:classical-freq-sectors}) (see also Sec.~\ref{sec:relObs-II}). Concretely, the degeneracy $\sigma$ is related to different frequency sectors of the theory if the constraint is quadratic in the momenta (see~\cite{Chataig:2019} for a discussion applied to the quantum relativistic particle and Sec.~\ref{sec:flrw} for an analysis of a quantum cosmological model). If only one multiplicity sector is defined, one may formally set $\hat{\Omega}_{\chi}^{\sigma} = \delta_{\sigma,0}\hat{\Omega}_{\chi}$ and drop the $\delta_{\sigma',\sigma}$ in~(\ref{eq:conditions-omega-2}). Equation~(\ref{eq:conditions-omega-1}) is a symmetric resolution of the identity in the physical Hilbert space and, therefore, it is a quantum version of the on-shell Faddeev-Popov resolution of the identity~(\ref{eq:FP-general-1}).

Classically, $\chi$ is admissible if~(\ref{eq:admissible-general}) is fulfilled, which implies that $\Delta_{\chi}^{-1}$ is well-defined and invertible [cf.~(\ref{eq:FP-general})] [if necessary, one needs to make the substitution~(\ref{eq:non-monotonic-subs})]. In the quantum theory, we shall consider the gauge condition $\hat{\chi}$ to be admissible if the invertible operator $\hat{\Omega}_{\chi}^{\sigma}$ can be determined from the orthonormality conditions~(\ref{eq:conditions-omega-1}) and~(\ref{eq:conditions-omega-2}) in a unique manner. In what follows, we assume that this is the case. 

We can now define quantum relational observables. Given any worldline-scalar operator $\hat{f}$, we define its associated relational observable relative to the gauge condition $\hat{\chi}$ as the on-shell operator\footnote{See~\cite{Chataig:2019} for a heuristic motivation.}
\begin{equation}\label{eq:relObs-quantum-1}
\hat{\Ob}_{(I)}[f|\chi = s]:= \pi\hbar\sum_{\sigma}\hat{\Omega}_{\chi}^{\sigma}[\hat{f},\hat{P}_{\chi = s}]_+\hat{\Omega}_{\chi}^{\sigma} \ ,
\end{equation}
where $[\cdot,\cdot]_+$ is the anticommutator,\footnote{Another factor ordering of the term $\hat{f}\hat{P}_{\chi = s}$ was suggested in~\cite{Chataig:2019}. In what follows, we will focus on the case in which $\hat{f}$ commutes with $\hat{\chi}$, such that the choice of ordering of the term $\hat{f}\hat{P}_{\chi = s}$ is irrelevant.}
\begin{equation}\notag
[\hat{f},\hat{P}_{\chi = s}]_+ = \hat{f}\hat{P}_{\chi = s}+\hat{P}_{\chi = s}\hat{f} \ ,
\end{equation}
and the subscript $(I)$ is included in order to distinguish~(\ref{eq:relObs-quantum-1}) from the subsequent definition~(\ref{eq:relObs-quantum-II}) to be analyzed in Sec.~\ref{sec:relObs-II}. Due to~(\ref{eq:conditions-omega-1}), we find the on-shell quantum analogue of~(\ref{eq:FP-general-0}),
\begin{equation}\label{eq:relObs-quantum-1-FP}
\hat{\Ob}_{(I)}[1|\chi = s] = \hat{P}_{E = 0} \ .
\end{equation}
Moreover, let us assume we can define the invariant operator [cf.~(\ref{eq:omega})]
\begin{equation}\notag
\left(\hat{\Delta}_{\chi}^{\sigma}\right)^{\frac{1}{2}}= \sum_{\bf k', k}\int_{\mathbb{R}}\D E\ \Omega^{\sigma}_{\chi}(E,{\bf k', k})\ket{E, \bf k'}\bra{E, \bf k} \ ,
\end{equation}
with $\Omega^{\sigma}_{\chi}(E = 0,{\bf k', k}) = \Omega^{\sigma}_{\chi}({\bf k', k})$. In this case, we can write 
\begin{equation}\label{eq:omega-delta-12}
\hat{\Omega}_{\chi}^{\sigma} = \hat{P}_{E = 0}\left(\hat{\Delta}_{\chi}^{\sigma}\right)^{\frac{1}{2}} \ ,
\end{equation}
such that~(\ref{eq:relObs-quantum-1}) becomes [cf.~(\ref{eq:general-on-shell-PoP})]
\begin{equation}\label{eq:relObs-quantum-2}
\hat{\Ob}_{(I)}[f|\chi = s]= 2\pi\hbar\hat{P}_{E = 0}\ \hat{\omega}_{(I)}[f|\chi = s]\ \hat{P}_{E = 0} \ ,
\end{equation}
where
\begin{equation}\label{eq:relObs-quantum-2-omega}
\hat{\omega}_{(I)}[f|\chi = s] :=\frac{1}{2}\sum_{\sigma}\left(\hat{\Delta}_{\chi}^{\sigma}\right)^{\frac{1}{2}}[\hat{f},\hat{P}_{\chi = s}]_+\left(\hat{\Delta}_{\chi}^{\sigma}\right)^{\frac{1}{2}} \ .
\end{equation}
Due to the relation between on-shell and general invariant operators given in~(\ref{eq:general-inv-quantum}) and~(\ref{eq:general-on-shell-PoP}), we conclude from~(\ref{eq:relObs-quantum-2}) and~(\ref{eq:relObs-quantum-2-omega}) that the quantum relational observable $\hat{\Ob}_{(I)}[f|\chi = s]$ may also be represented as the invariant
\begin{equation}\label{eq:relObs-quantum-3}
\hat{\Ob}_{(I)}[f|\chi = s] = \int_{-\infty}^{\infty}\D\tau\ \e^{\frac\I\hbar\tau\hat{C}}\ \hat{\omega}_{(I)}[f|\chi = s]\ \e^{-\frac\I\hbar\tau\hat{C}} \ ,
\end{equation}
which is a quantum version of~(\ref{eq:relObs-general}). We note that $\hat{P}_{\chi = s}$ is the operator version of $\delta(\chi(q(\tau),p(\tau))-s)$ and $\left(\hat{\Delta}_{\chi}^{\sigma}\right)^{\frac{1}{2}}$ corresponds to the square root of the classical quantity $\Delta_{\chi}$ in~(\ref{eq:relObs-general}).

Finally, let us assume that $\hat{f}$ commutes with the gauge condition $\hat{\chi}$; i.e., 
\begin{equation}\label{eq:f-commutes-chi}
\begin{aligned}
\hat{f} &:= \sum_{\bf n',n}\int_{\mathbb{R}}\D\chi\ \tilde{f}(\chi,{\bf n',n}) \ket{\chi,\bf n'}\bra{\chi,\bf n}\\
&=\sum_{\bf m}\int_{\mathbb{R}}\D\chi\ f(\chi,{\bf m}) \ket{\chi,\bf m}\bra{\chi,\bf m} \ ,
\end{aligned}
\end{equation}
where $\ket{\chi,\bf m}$ is the simultaneous complete orthonormal eigenbasis of $\hat{\chi}$ and $\hat{f}$. In this case, the on-shell relational observable $\hat{\Ob}_{(I)}[f|\chi = s]$ given in~(\ref{eq:relObs-quantum-1}) has the eigenstates
\begin{equation}\label{eq:eigenstate-relObs}
\ket{\sigma,{\bf m};s} := \sqrt{2\pi\hbar}\hat{\Omega}_{\chi}^{\sigma}\ket{\chi = s,\bf m} \ , 
\end{equation}
which form a complete orthonormal system with respect to the induced inner product due to~(\ref{eq:conditions-omega-1}),~(\ref{eq:conditions-omega-2}) and the fact that the bases $\ket{\chi,{\bf n}}$ and $\ket{\chi,{\bf m}}$ form a complete orthonormal system with respect to the auxiliary inner product. Indeed, due to~(\ref{eq:conditions-omega-1}) and~(\ref{eq:conditions-omega-2}), we may write
\begin{equation}\label{eq:SONC-eigenstates-relObs}
\begin{aligned}
&\sum_{\sigma}\sum_{\bf m}\ket{\sigma,{\bf m};s}\bra{\sigma,{\bf m};s} = \hat{P}_{E = 0} \ ,\\
&\left(\sigma',{\bf m'};s|\sigma,{\bf m};s\right) = \delta_{\sigma',\sigma}\delta({\bf m',m}) \ .
\end{aligned}
\end{equation}
We note that the observables~(\ref{eq:relObs-quantum-1}) may be interpreted as gauge-fixed\footnote{As was stressed in~\cite{Chataig:2019} (see also references therein), relational observables are invariant extensions of gauge-fixed quantities. Since we are only interested in on-shell invariants, we will take the terms ``gauge-fixed quantity'' and ``invariant extension of a gauge-fixed quantity'' to be synonyms. Indeed, one may consider that the quantities written in a particular gauge are gauge-invariant precisely because it is in principle possible to invariantly extend them~\cite{HT:book} if the canonical gauge condition is well defined [cf.~(\ref{eq:admissible-general})]. For this reason, in a slight abuse of terminology, we refer to the on-shell relational observables as gauge-fixed operators. In the same way, we refer to their equations of motion found for a particular case in Sec.~\ref{sec:relObs-II} as the gauge-fixed Heisenberg equations and to the overlap of their eigenstates as the gauge-fixed propagator [cf. Sec.~\ref{sec:gf-propagator}].} Heisenberg-picture operators, while the states~(\ref{eq:eigenstate-relObs}) can be seen as their Heisenberg-picture eigenstates [see also Eq.~(\ref{eq:gf-Heisenberg}) and the discussion in Sec.~\ref{sec:gf-propagator}].

An earlier version of this construction was extensively analyzed in~\cite{Chataig:2019}, where its relation to the classical expression~(\ref{eq:relObs-general}) and its evolution were examined for the examples of a free relativistic particle and a vacuum Bianchi I model.

\subsubsection{\label{sec:relObs-II}Quantum relational observables II}
We now analyze an alternative formulation of the construction of quantum relational observables for the particular case in which the canonical momentum conjugate to the quantum gauge condition is invariant, i.e., it commutes with $\hat{C}$. To consider the most general instance of this case, instead of choosing a quantum gauge condition $\hat{\chi}$, we first choose an invariant Hamiltonian and subsequently derive what is the canonical gauge to be used.

Since we assume that the eigenstates of the constraint operator are degenerate (which is equivalent to assuming the reduced phase space is not trivial), it is possible to define arbitrary invariant operators as follows:
\begin{equation}
\hat{H}_{\chi} := \sum_{\bf k',k}\int_{\mathbb{R}}\D E\ H_{\chi}(E,{\bf k',k})\ket{E,{\bf k'}}\bra{E,{\bf k}} \ ,
\end{equation}
where $H_{\chi}(E,{\bf k',k}) = \overline{H_{\chi}(E,{\bf k,k'})}$. Equivalently, one could choose an invariant of the form~(\ref{eq:general-inv-quantum}). Analogously to $\hat{C}$, we require that the spectrum of $\hat{H}_{\chi}$ is continuous and ranges over $\mathbb{R}$. If such an $\hat{H}_{\chi}$ can be defined, it commutes with $\hat{C}$, and thus they are simultaneously diagonalizable. Let us denote the simultaneous complete orthonormal eigenbasis as $\ket{h,{\bf n}}$ and write
\begin{equation}\label{eq:function-eigenvalues-Ch}
\begin{aligned}
\hat{C}\ket{h,{\bf n}} &= C(h,{\bf n})\ket{h,{\bf n}} \ , \\
\hat{H}_{\chi}\ket{h,{\bf n}} &= h\ket{h,{\bf n}} \ ,
\end{aligned}
\end{equation}
where $C(h,{\bf n})$ is a real function. More precisely, Eq.~(\ref{eq:function-eigenvalues-Ch}) corresponds to the situation in which one finds a complete set of commuting invariants, the eigenstates of which are labeled by $h,{\bf n}$, in terms of which the eigenvalues of constraint operator are expressed. We will see a concrete example in Sec.~\ref{sec:flrw}. If we take $\hat{H}_{\chi}$ to be a Hamiltonian, we can find the corresponding time operator (gauge condition) $\hat{\chi}$ as follows.\footnote{We note that the Pauli theorem~\cite{QRef5,Pauli} can be avoided due to the assumption that $h\in\mathbb{R}$. However, the on-shell, physical Hamiltonian will possibly be bounded from below or above in each multiplicity sector (see below) due to the restriction of $\hat{H}_{\chi}$ to the physical Hilbert space.} Consider the states
\begin{equation}\label{eq:chi-cov}
\ket{\chi,{{\bf n}}} := \int_{\mathbb{R}}\frac{\D h}{\sqrt{2\pi\hbar}}\ \e^{\frac{\I}{\hbar}h\chi}\ket{h,{\bf n}}\ , 
\end{equation}
which satisfy the orthogonality condition
\begin{equation}
\begin{aligned}
\braket{\chi',{{\bf n}'}|\chi,{{\bf n}}} &= \int_{\mathbb{R}}\frac{\D h}{2\pi\hbar}\e^{\frac{\I}{\hbar}h(\chi-\chi')}\delta({{\bf n}',{\bf n}})\\
&=\delta(\chi'-\chi)\delta({{\bf n}',{\bf n}})\ ,
\end{aligned}
\end{equation}
and the completeness relation
\begin{align*}
\sum_{{\bf n}}\int_{\mathbb{R}}\D\chi\ \ket{\chi,{{\bf n}}}\bra{\chi,{{\bf n}}}=\sum_{{\bf n}}\int_{\mathbb{R}}\D h\ \ket{h,{{\bf n}}}\bra{h,{{\bf n}}} = \hat{1} \ .
\end{align*}
Due to~(\ref{eq:chi-cov}), the operator
\begin{equation}
\hat{\chi}:=\sum_{{\bf n}}\int_{\mathbb{R}}\D\chi\ \chi \ket{\chi,{{\bf n}}}\bra{\chi,{{\bf n}}} 
\end{equation}
satisfies the ``covariance'' property
\begin{equation}\label{eq:cov-of-chi}
\e^{\frac{\I}{\hbar}\hat{H}_{\chi}s}\hat{\chi}\e^{-\frac{\I}{\hbar}\hat{H}_{\chi}s} = \hat{\chi}-s\hat{1}.
\end{equation}
For this reason, $\hat{\chi}$ is a ``covariant time operator''~\cite{Hoehn:Trinity,QRef4,QRef5} associated with the Hamiltonian $\hat{H}_{\chi}$. By differentiating~(\ref{eq:cov-of-chi}) with respect to $s$ and setting $s = 0$~\cite{Hoehn:Trinity}, we find the formal commutation relation\footnote{This is the correct (formal) commutation relation between a (reduced phase-space) Hamiltonian and the associated time operator, i.e., the Hamiltonian plays the role of the opposite of the canonical momentum conjugate to $\hat{\chi}$ [cf.~(\ref{eq:classical-freq-sectors}) and~(\ref{eq:on-shell-action-general-2})].}
\begin{equation}
[\hat{\chi},\hat{H}_{\chi}] = -\I\hbar\hat{1} \ .
\end{equation}
This construction has thus far been with respect to the auxiliary Hilbert space. We are now in a position to construct on-shell observables in the physical Hilbert space.

First, in analogy to the classical equation~(\ref{eq:classical-freq-sectors}), we assume that, for a certain interval of values of $E$ that contains $E = 0$, the equation
\begin{equation}\label{eq:ChnE}
C(h,{\bf n}) = E
\end{equation}
has the real solutions
\begin{equation}\label{eq:quantum-freq-sectors}
h = -H_{\chi}^{\sigma}(E,{{\bf n}}) \ ,
\end{equation}
where $\sigma$ denotes a possible discrete degeneracy; i.e., if $\sigma'\neq\sigma$, then $H_{\chi}^{\sigma'}(E,{{\bf n}})\neq H_{\chi}^{\sigma}(E,{{\bf n}})$ for all allowed values of $E,{\bf n}$. We also define
\begin{equation}\label{eq:on-shell-Esigman}
\frac{1}{\mathcal{N}}\ket{E, \sigma, {\bf n}} := \ket{h, {\bf n}}_{h = -H_{\chi}^{\sigma}(E,{{\bf n}})} \ ,
\end{equation}
where $\mathcal{N}\equiv\mathcal{N}(E,\sigma,{\bf n})$ is a normalization factor. Second, we note that the induced inner product of the states $\ket{E, \sigma, {\bf n}}$ can be computed from the auxiliary overlap of $\ket{h,{\bf n}}$,
\begin{equation}\label{eq:auxiliary-overlap-relObs2}
\begin{aligned}
\braket{h',{\bf n}'|h,{\bf n}}&=\delta(h'-h)\delta({\bf n}',{\bf n})\\
&=\delta(E'-E)\frac{(E,\sigma',{\bf n}'|E,\sigma,{\bf n})}{\mathcal{N}^2} \ ,
\end{aligned}
\end{equation}
where the induced product reads
\begin{equation}\label{eq:induced-overlap-relObs2}
(E,\sigma',{\bf n}'|E,\sigma,{\bf n}):=\mathcal{N}^2\delta_{\sigma',\sigma}\delta({\bf n}',{\bf n})\left|\frac{\partial C}{\partial h}\right|_{h = -H_{\chi}^{\sigma}(E,{{\bf n}})} \ .
\end{equation}
The Kronecker delta $\delta_{\sigma',\sigma}$ is included in~(\ref{eq:induced-overlap-relObs2}) because $\delta(h'-h)$ is zero if $h', h$ belong to different $\sigma$-sectors [cf. discussion after~(\ref{eq:quantum-freq-sectors})]. We normalize the states $\ket{E, \sigma,{\bf n}}$ with respect to the induced inner product by setting
\begin{equation}\label{eq:induced-overlap-relObs2-normalization}
\mathcal{N} = \left|\frac{\partial C}{\partial h}\right|_{h = -H_{\chi}^{\sigma}(E,{{\bf n}})}^{-\frac{1}{2}} \ .
\end{equation}
For convenience, we will also denote on-shell states by the abbreviated notation $\ket{\sigma,{\bf n}}:=\ket{E = 0, \sigma,{\bf n}}$. We can now define the (improper) projectors 
\begin{align}
\hat{P}_{E}^{\sigma}&:=\sum_{\bf n}\ket{E,\sigma,{\bf n}}\bra{E,\sigma,{\bf n}} \label{eq:projectors-relObs2}\ ,\\
\hat{P}_{E}&:= \sum_{\sigma}\hat{P}_{E}^{\sigma} \ , \label{eq:PE-sum-PEsigma}
\end{align}
which, due to~(\ref{eq:auxiliary-overlap-relObs2}) and~(\ref{eq:induced-overlap-relObs2}), satisfy $\hat{P}_{E'}^{\sigma'}\hat{P}_{E}^{\sigma} = \delta_{\sigma',\sigma}\delta(E'-E)\hat{P}_{E}^{\sigma}$. The quantum analogue of~(\ref{eq:FP-general}) is defined to be the self-adjoint on-shell operator~[cf.~(\ref{eq:general-on-shell-PoP})]
\begin{equation}\label{eq:FP-general-quantum}
\left(\hat{\Omega}_{\chi}^{\sigma}\right)^{-2}:= 2\pi\hbar\hat{P}_{E=0}^{\sigma}\hat{P}_{\chi = s}\hat{P}_{E=0}^{\sigma} \ ,
\end{equation}
which will be referred to as the inverse Faddeev-Popov operator for the $\sigma$-sector. To find its matrix elements, we compute the overlap [cf.~(\ref{eq:chi-cov})]
\begin{equation}
\braket{\sigma,{\bf n}'|\chi,{\bf n}} = \left(\Omega_{\chi}^{\sigma}\right)^{-1}\frac{\delta({\bf n',n})}{\sqrt{2\pi\hbar}}\left.\e^{\frac{\I}{\hbar}h\chi}\right|_{h = -H_{\chi}^{\sigma}(0,{{\bf n}})} \ ,
\end{equation}
where
\begin{equation}\label{eq:matrix-element-FP}
\Omega_{\chi}^{\sigma} \equiv \Omega_{\chi}^{\sigma}({\bf n}):= \left|\frac{\partial C}{\partial h}\right|_{h = -H_{\chi}^{\sigma}(0,{{\bf n}})}^{\frac{1}{2}} \ .
\end{equation}
We thus obtain
\begin{equation}\label{eq:FP-general-quantum-2}
\left(\hat{\Omega}_{\chi}^{\sigma}\right)^{-2}:= \sum_{\bf n}\left(\Omega_{\chi}^{\sigma}({\bf n})\right)^{-2}\ket{\sigma,{\bf n}}\bra{\sigma,{\bf n}} \ .
\end{equation}
If we now define, more generally, the operators
\begin{equation}\label{eq:FP-general-quantum-powers}
\left(\hat{\Omega}_{\chi}^{\sigma}\right)^{\rho}:= \sum_{\bf n}\left(\Omega_{\chi}^{\sigma}({\bf n})\right)^{\rho}\ket{\sigma,{\bf n}}\bra{\sigma,{\bf n}} \ ,
\end{equation}
and if we use~(\ref{eq:induced-overlap-relObs2}) together with~(\ref{eq:induced-overlap-relObs2-normalization}), we find
\begin{align}
\hat{P}_{E = 0}^{\sigma}\bullet\left(\hat{\Omega}_{\chi}^{\sigma}\right)^{\rho} &= \left(\hat{\Omega}_{\chi}^{\sigma}\right)^{\rho}\bullet\hat{P}_{E = 0}^{\sigma} = \left(\hat{\Omega}_{\chi}^{\sigma}\right)^{\rho} \ ,\label{eq:PDelta-FP}\\
\hat{\Omega}_{\chi}^{\sigma}\bullet\left(\hat{\Omega}_{\chi}^{\sigma}\right)^{-1} &= \left(\hat{\Omega}_{\chi}^{\sigma}\right)^{-1}\bullet\hat{\Omega}_{\chi}^{\sigma} = \hat{P}_{E = 0}^{\sigma} \ .\label{eq:Delta-Inversion-FP}
\end{align}
Thus, we obtain from~(\ref{eq:FP-general-quantum}),~(\ref{eq:PDelta-FP}) and~(\ref{eq:Delta-Inversion-FP}) the Faddeev-Popov resolution of the identity in the $\sigma$-sector of the physical Hilbert space,
\begin{equation}\label{eq:FP-identity-quantum}
\begin{aligned}
\hat{P}_{E = 0}^{\sigma} =& 2\pi\hbar\left(\hat{\Omega}_{\chi}^{\sigma}\right)^{2}\hat{P}_{\chi = s}\hat{P}_{E=0}^{\sigma} \\
=&2\pi\hbar\hat{P}_{E=0}^{\sigma}\hat{P}_{\chi = s}\left(\hat{\Omega}_{\chi}^{\sigma}\right)^{2} \ ,
\end{aligned}
\end{equation}
which can also be written symmetrically as
\begin{equation}\label{eq:FP-identity-quantum-sym}
\hat{P}_{E = 0}^{\sigma} = 2\pi\hbar\hat{\Omega}_{\chi}^{\sigma}\hat{P}_{\chi = s}\hat{\Omega}_{\chi}^{\sigma} \ .
\end{equation}
Equations~(\ref{eq:FP-identity-quantum}) and~(\ref{eq:FP-identity-quantum-sym}) are quantum analogues of~(\ref{eq:FP-general-1}) and correspond to different factor ordering choices for the quantum relational observables. Indeed, Eq.~(\ref{eq:FP-identity-quantum-sym}) is equivalent to~(\ref{eq:conditions-omega-1}), from which we can define the relational observable $\hat{\Ob}_{(I)}[f|\chi = s]$ given in~(\ref{eq:relObs-quantum-1}). From~(\ref{eq:FP-identity-quantum}), we see that another possible factor ordering is\footnote{The abbreviation ``h.c.'' stands for ``Hermitian conjugate''.}
\begin{equation}\label{eq:relObs-quantum-II}
\hat{\Ob}_{(II)}[f|\chi = s]:=\pi\hbar\sum_{\sigma}\left(\hat{\Omega}_{\chi}^{\sigma}\right)^2\hat{P}_{\chi = s}\hat{f}\hat{P}_{E=0}^{\sigma}+\text{h.c.} \ ;
\end{equation}
i.e., $\hat{\Ob}_{(II)}[f|\chi = s]$ is another possible definition of the on-shell quantum relational observable associated with a worldline scalar $\hat{f}$ relative to the gauge condition $\hat{\chi}$.  By using~(\ref{eq:PE-sum-PEsigma}) and~(\ref{eq:FP-identity-quantum}), we find
\begin{equation}
\hat{\Ob}_{(II)}[1|\chi = s] = \hat{P}_{E=0} \ ,
\end{equation}
just as $\hat{\Ob}_{(I)}[1|\chi = s] = \hat{P}_{E = 0}$ in~(\ref{eq:relObs-quantum-1-FP}). The definition~(\ref{eq:relObs-quantum-II}) also has the attractive feature that the observable associated with an invariant of the form
\begin{equation}\label{eq:sigma-invariant-f}
\hat{f} := \sum_{\sigma}\sum_{\bf n',n}\int\D E\ f^{\sigma}(E,{\bf n', n})\ket{E,\sigma,{\bf n'}}\bra{E,\sigma,{\bf n}}
\end{equation}
is just $\hat{f}\hat{P}_{E = 0}$, due to~(\ref{eq:FP-identity-quantum}) [this is true provided $\hat{f}$ in~(\ref{eq:sigma-invariant-f}) is well-defined, i.e., that the integral over $E$ can be performed]. We will, however, focus on operators $\hat{f}$ that commute with the gauge condition~[cf.~(\ref{eq:f-commutes-chi})]. In this case, we can find a dynamical equation for $\hat{\Ob}_{(I,II)}[f|\chi = s]$ as follows. Due to~(\ref{eq:function-eigenvalues-Ch}), we obtain
\begin{equation}
\hat{H}_{\chi}\ket{\sigma,{\bf n}} = -H_{\chi}^{\sigma}(0,{\bf n})\ket{\sigma,{\bf n}} \ ,
\end{equation}
which, together with~(\ref{eq:FP-general-quantum-powers}), implies
\begin{equation}\label{eq:delta12-commutes-H}
\left[\left(\hat{\Omega}_{\chi}^{\sigma}\right)^{\rho},\hat{H}_{\chi}\right] = 0 \ .
\end{equation}
Moreover, Eqs.~(\ref{eq:f-commutes-chi}) and~(\ref{eq:chi-cov}) yield
\begin{equation}\label{eq:f-derivative-s}
\begin{aligned}
\I\hbar\frac{\D}{\D s}\left(\hat{f}\hat{P}_{\chi=s}\right) =& \sum_{\bf m}\I\hbar\frac{\partial f}{\partial s}\ket{\chi = s,{\bf m}}\bra{\chi = s,{\bf m}}\\
&+[\hat{f}\hat{P}_{\chi =s},\hat{H}_{\chi}] \ .
\end{aligned}
\end{equation}
Thus, using~(\ref{eq:delta12-commutes-H}) and~(\ref{eq:f-derivative-s}), we may write the derivative of $\hat{\Ob}_{(I,II)}[f|\chi = s]$ as
\begin{equation}\label{eq:gf-Heisenberg}
\frac{\D}{\D s}\hat{\Ob}[f|\chi = s] = \hat{\Ob}\left[\left.\frac{\partial f}{\partial s}\right|\chi = s\right]+\frac{1}{\I\hbar}[\hat{\Ob}[f|\chi = s],\hat{H}_{\chi}] \ ,
\end{equation}
where we have suppressed the $(I,II)$ subscripts for brevity, as~(\ref{eq:gf-Heisenberg}) holds for both definitions. We interpret~(\ref{eq:gf-Heisenberg}) as the gauge-fixed Heisenberg equation of motion\footnote{In the derivation of~(\ref{eq:gf-Heisenberg}), we assumed that the Hamiltonian $\hat{H}_{\chi}$ does not depend on the gauge-fixed time parameter $s$.} for the relational observables.

\subsubsection{\label{sec:gf-propagator}Gauge-fixed propagator}
The quantum relational observables~(\ref{eq:relObs-quantum-1}) and~(\ref{eq:relObs-quantum-II}) are formally symmetric with respect to the auxiliary inner product and they commute with the constraint operator $\hat{C}$. Therefore, they are formally symmetric with respect to the induced inner product~(\ref{eq:induced-IP}). If we furthermore assume that they are self-adjoint, one may find a complete orthonormal system of eigenstates of the observables. In this case, every on-shell state can thus be expanded into the bases of eigenstates of relational observables for a fixed value of $s$. The physical interpretation of this fact will be discussed in Secs.~\ref{sec:switch} and~\ref{sec:invext}.

Let us consider for definiteness the eigenstates of $\hat{\Ob}_{(I)}[f|\chi = s]$ given in~(\ref{eq:eigenstate-relObs}). By construction, they form a complete orthonormal system in the physical Hilbert space due to the orthonormality conditions~(\ref{eq:conditions-omega-1}) and~(\ref{eq:conditions-omega-2}). As was remarked at the end of Sec.~\ref{sec:relObs-I}, such states are the eigenstates of the gauge-fixed Heisenberg-picture operators. For a fixed value $s = s_0$, we can write any on-shell state as
\begin{equation}
\ket{\Psi} = \sum_{\sigma}\sum_{\bf m}\Psi_{\sigma}({\bf m})\ket{\sigma,{\bf m};s_0} \ .
\end{equation}
We define $\left(\sigma',{\bf m'};s'|\sigma,{\bf m},s\right)$ to be the gauge-fixed propagator, which encodes the evolution with respect to the gauge-fixed time parameter $s$. Due to the orthonormality of the states $\ket{\sigma,{\bf m};s}$ for all values of $s$ [cf.~(\ref{eq:SONC-eigenstates-relObs})], we have
\begin{equation}\label{eq:gf-propagator-limit}
\lim_{s'\to s}\left(\sigma',{\bf m'};s'|\sigma,{\bf m},s\right) = \delta_{\sigma',\sigma}\delta({\bf m',m}) \ .
\end{equation}
To see that this evolution corresponds to a unitary transformation in the physical Hilbert space (regardless of whether $\hat{p}_{\chi}$ is an invariant, as considered in Sec.~\ref{sec:relObs-II}), we consider the (gauge-fixed Schr\"{o}dinger-picture) state
\begin{equation}
\ket{\Psi;s}=\sum_{\sigma}\sum_{\bf m}\Psi_{\sigma}({\bf m};s)\ket{\sigma,{\bf m};s_0} \ ,
\end{equation}
where
\begin{equation}
\Psi_{\sigma}({\bf m};s) := \sum_{\sigma_0}\sum_{{\bf m}_0}\left(\sigma,{\bf m};s|\sigma_0,{\bf m}_0;s_0\right)\Psi_{\sigma_0}({\bf m}_0) \ .
\end{equation}
Due to the completeness of the states $\ket{\sigma,{\bf m};s}$ for all values of $s$ [cf.~(\ref{eq:SONC-eigenstates-relObs})], we thus obtain
\begin{align*}
&\sum_{\sigma}\sum_{\bf m}|\Psi_{\sigma}({\bf m};s)|^2\\
&=\sum_{\sigma_0,\sigma_0',\sigma}\sum_{{\bf m}_0,{\bf m'}_0,{\bf m}}\overline{\Psi}_{\sigma'_0}({\bf m'}_0)\left(\sigma'_0,{\bf m'}_0;s_0|\sigma,{\bf m};s\right)\\
&\times\left(\sigma,{\bf m};s|\sigma_0,{\bf m}_0;s_0\right)\Psi_{\sigma_0}({\bf m}_0)\\
&=\sum_{\sigma_0,\sigma_0'}\sum_{{\bf m}_0,{\bf m'}_0}\overline{\Psi}_{\sigma'_0}({\bf m'}_0)\left(\sigma'_0,{\bf m'}_0;s_0|\sigma_0,{\bf m}_0;s_0\right)\Psi_{\sigma_0}({\bf m}_0)\\
&= \sum_{\sigma_0}\sum_{{\bf m}_0}|\Psi_{\sigma_0}({\bf m}_0)|^2\ .
\end{align*}
Therefore, the norm of the on-shell states is conserved,
\begin{equation}\label{eq:gf-prop-unitary}
(\Psi;s|\Psi;s) = (\Psi|\Psi) \ .
\end{equation}

\subsubsection{\label{sec:QRef}Quantum reference frames}
What is the physical interpretation of the quantum observables~(\ref{eq:relObs-quantum-1}) or~(\ref{eq:relObs-quantum-II})? As previously stated, the classical relational observable~(\ref{eq:relObs-general}) represents the value of $f(q(\tau),p(\tau))$ in relation to the value $s$ of $\chi(q(\tau),p(\tau))$, i.e., it is the value of $f(q(\tau),p(\tau))$ when $\chi(q(\tau),p(\tau)) = s$. Do $\hat{\Ob}_{(I,II)}[f|\chi = s]$ represent $\hat{f}$ ``in relation to'' the eigenvalue $s$ of $\hat{\chi}$? If so, in what sense? The formalism of~\cite{Chataig:2019} and the earlier treatments~\cite{Rovelli:1990-1,Rovelli:1990-2,Rovelli:1991,Tambornino:2012} do not seem to provide a satisfactory answer at the conceptual level.

The structural similarity of~(\ref{eq:relObs-quantum-3}) with its classical counterpart guarantees that certain properties of the classical observables are translated into the quantum theory. For instance, as shown in~\cite{Chataig:2019} and as was discussed in Sec.~\ref{sec:relObs-II}, the quantum observables $\hat{\Ob}_{(I,II)}[f|\chi = s]$ obey gauge-fixed Heisenberg equations which are the quantum analogues of the classical reduced phase-space equations. This is true provided one works with a gauge condition that is conjugate to an invariant Hamiltonian operator. Furthermore, as shown in Sec.~\ref{sec:gf-propagator}, the gauge-fixed evolution encoded in the gauge-fixed propagator is a unitary transformation in the physical Hilbert space.

As remarked in Sec.~\ref{sec:classical-general}, the reduced phase-space equations can be thought of as the field equations in a particular reference frame, in which the time coordinate $s$ is defined by the level sets of $\chi(q(\tau),p(\tau))$. In this way, we may also consider that the gauge-fixed Heisenberg equations~(\ref{eq:gf-Heisenberg}) or the gauge-fixed propagator~(\ref{eq:gf-propagator-limit}), which describe the quantum dynamics with respect to the c-number $s$, are related to (or can be used to define) a notion of ``quantum reference frames''. Given a time-reparametrization invariant quantum system, we consider that a choice of reference frame corresponds to a partitioning of the system into a reference (the gauge condition $\hat{\chi}$) and a subsystem, which is the set of all worldline tensors $\hat{f}$ that commute with $\hat{\chi}$ (see also the discussion in~\cite{QRef3,QRef4}). The quantum relational observables [constructed as in~(\ref{eq:relObs-quantum-1}) or~(\ref{eq:relObs-quantum-II})] thus describe the evolution of the subsystem in the reference frame defined by $\hat{\chi}$. A similar point of view was expressed in~\cite{Chataig:2019}, where it was argued that the quantum relational observables capture all the relational content of the theory. We do not dispute this view in the present article, but we wish to further understand (in a precise yet pragmatic way) in what sense the theory of quantum relational observables is indeed ``relational''. In Sec.~\ref{sec:CP-general}, we suggest that a clearer physical interpretation of such observables can be obtained if we make use of conditional probabilities.

It is worthwhile to mention that a great deal of effort has been devoted to the precise definition and study of quantum reference frames and ``relational quantum clocks'' (see~\cite{Hoehn:2018-1,Hoehn:2018-2,Hoehn:Trinity,Hoehn:2019,QRef1,QRef3,QRef4,QRef5} and references therein). In particular, the authors of~\cite{Hoehn:2018-1,Hoehn:2018-2,Hoehn:Trinity,Hoehn:2019} establish a formalism to relate different choices of quantum reference frames and discuss under which circumstances these choices can be related to a direct quantization of the reduced phase-space equations that can be derived from the action~(\ref{eq:on-shell-action-general}). In~\cite{Hoehn:2018-2}, it is discussed in detail how one may map the physical Hilbert space of on-shell states (as presented in Sec.~\ref{sec:QT:Hilbert} of the present article) to the different Hilbert spaces associated with the various quantizations of the reduced phase-space equations for different choices of $\chi$ in~(\ref{eq:on-shell-action-general}). We note that an early reference with a similar investigation is~\cite{Barvinsky}, in which Barvinsky relates the transition amplitudes of on-shell states (expressed as path integrals) with the reduced phase-space path integrals written in terms of the action~(\ref{eq:on-shell-action-general}). Thus, the work of~\cite{Barvinsky} can be seen as a possible formalism to relate different choices of reference frames (defined from the reduced phase-space path integrals) with the physical Hilbert space of on-shell states. In~\cite{Barvinsky}, the canonical (operator-based) theory associated with the various path integrals is also discussed at the semiclassical (one-loop) order.

The (in principle) different approaches of~\cite{Hoehn:Trinity,QRef3,QRef4,Barvinsky} point to the conclusion that different choices of reference frames (different choices of $\chi$) can be described in a single space, which is the physical Hilbert space of on-shell states (variously referred to as the ``Dirac-Wheeler-DeWitt formulation'' in~\cite{Barvinsky} and as the ``perspective-neutral'' or ``reference-system-neutral'' framework in~\cite{Hoehn:2018-2}). This is also the case in the formalism here presented. In Sec.~\ref{sec:switch}, we use the quantum relational observables and the gauge-fixed propagator here defined to argue that a change in quantum reference frames indeed corresponds to a change of basis in the physical Hilbert space. Moreover, in the recent article~\cite{Hoehn:Trinity}, it was shown that the formalism of~\cite{Hoehn:2018-2} was equivalent to the use of conditional probabilities for a simple class of models. In the same spirit, in Sec.~\ref{sec:CP-general}, we argue that the use of conditional probabilities can elucidate the physical interpretation of the quantum relational observables here defined and their connection to quantum reference frames without the restriction of~\cite{Barvinsky} to a perturbative treatment in $\hbar$. Finally, in Sec.~\ref{sec:PW}, we comment on how the formalism here presented can be related to the relativization maps and $G$-twirl operations employed in~\cite{Hoehn:Trinity,QRef1,QRef3,QRef4,QRef5}.

\subsubsection{\label{sec:switch}Switching reference frames}
As we have seen in Sec.~\ref{sec:gf-propagator}, one may expand any on-shell state $\ket{\Psi}$ into the bases of eigenstates of relational observables. From the discussion in Sec.~\ref{sec:QRef}, we thus interpret the overlap $(\sigma,{\bf m};s|\Psi)$ as the representation of the on-shell state in the quantum reference frame defined by the gauge condition $\hat{\chi}$. We will see in Sec.~\ref{sec:CP-general} that this overlap may also be interpreted in terms of conditional probabilities.

Can we switch reference frames in this formalism? Yes, the switch corresponds to a change of basis in the physical Hilbert space.\footnote{This is reminiscent of the assertion made in~\cite{HT:Vergara} (see also~\cite{HT:book}) that the difference between the reduced phase-space path integral and the path integral related to the transition amplitude of on-shell states corresponds to the difference between the kernels of an operator in different representations.} Let $\hat{\chi}_1$ and $\hat{\chi}_2$ be two admissible gauge conditions and $\ket{\sigma_1,{\bf m}_1;\chi_1}, \ket{\sigma_2,{\bf m}_2;\chi_2}$ be the corresponding complete orthonormal systems of eigenstates of relational observables [cf.~Sec.~\ref{sec:relObs-I}]. To switch from the representation of an on-shell state $\ket{\Psi}$ in the reference frame defined by $\hat{\chi}_1$ to the one defined by $\hat{\chi}_2$, we insert the corresponding resolution of the identity,
\begin{align*}
&(\sigma_1,{\bf m}_1;\chi_1|\Psi)\\
&= \sum_{\sigma_2}\sum_{{\bf m}_2}(\sigma_1,{\bf m}_1;\chi_1|\sigma_2,{\bf m}_2;\chi_2)(\sigma_2,{\bf m}_2;\chi_2|\Psi) \ .
\end{align*}
We note that the gauge-fixed propagator [cf.~Sec.~\ref{sec:gf-propagator}] is a particular case of the change of basis matrix $(\sigma_1,{\bf m}_1;\chi_1|\sigma_2,{\bf m}_2;\chi_2)$, in which instead of switching reference frames, one simply switches the value $s$ of the gauge condition.

\subsubsection{\label{sec:CP-general}Conditional probabilities}
What is the meaning of the gauge-fixed time parameter $s$ in the quantum theory? As we have argued, it is the time parameter in a particular reference frame, but how is it related to observations? Classically, $s$ is a value of $\chi(q(\tau),p(\tau))$. Once we know $\chi(q(\tau),p(\tau)) = s$ (via an observation), we can predict, in a diffeomorphism-invariant fashion, what is the value of $f(q(\tau),p(\tau))$ by computing $\Ob[f|\chi = s]$. Thus, classical relational observables are conditional quantities, in the sense that they yield predictions based on a certain condition (the observed value of $\chi(q(\tau),p(\tau))$). It is thus reasonable to consider that what one should compute in the quantum theory is the probability of observing a certain (eigen)value of $\hat{f}$, given that $\hat{\chi}$ is in the state $\ket{\chi = s,\bf n}$. In other words, we should consider conditional probabilities defined from the on-shell states~(\ref{eq:on-shell-state-general}).

There is an extensive literature on the use of conditional probabilities in time-reparametrization invariant mechanical systems (see~\cite{PW1,PW2,PW3,PW4,PW5,PW6,PW7,Dolby:2004,Hoehn:Trinity} and references therein; see also Sec.~\ref{sec:PW}). Most articles focus on models in which the constraint $\hat{C}$ can be written as
\begin{equation}\label{eq:PW-constraint}
    \hat{C}\equiv C(\hat{q},\hat{p}) = C_{(1)}(\hat{q}^1,\hat{p}_1)+C_{(>1)}(\hat{q},\hat{p}) \ ,
\end{equation}
where $C_{(>1)}(\hat{q},\hat{p})$ only depends on $\hat{q}^i,\hat{p}_i$ for $i > 1$. For this form of the constraint operator, a variable canonically conjugate to $C_{(1)}(\hat{q}^1,\hat{p}_1)$ is usually chosen to play the role of time (the quantity $\hat{\chi}$ in the present article). Furthermore, the connection between relational observables and conditional probabilities was analyzed in~\cite{Hoehn:Trinity} (see also~\cite{Dolby:2004}) for models with constraints of the form~(\ref{eq:PW-constraint}) and gauge conditions $\hat{\chi}$ (formally) conjugate to $C_{(1)}(\hat{q}^1,\hat{p}_1)$.

As we wish to keep the discussion as general as possible, we do not restrict the constraint to the form~(\ref{eq:PW-constraint}), and we allow a gauge condition $\hat{\chi}$ that is admissible according to the criterion discussed in Sec.~\ref{sec:relObs-I}, but not necessarily conjugate to $C_{(1)}(\hat{q}^1,\hat{p}_1)$. In this framework, we establish a connection between conditional probabilities and quantum relational observables, which clarifies their physical meaning. We will comment on the relation of this formalism to previous approaches to conditional probabilities in Secs.~\ref{sec:Other-CP} and~\ref{sec:PW}.

For a general on-shell state $\ket{\Psi}$ defined as in~(\ref{eq:on-shell-state-general}), we postulate that
\begin{equation}\label{eq:CP}
    p_{\Psi}({\bf m}|\chi = s) = \frac{\left|\left<\chi = s, {\bf m}|\Psi\right>\right|^2}{\sum_{\bf m}\left|\left<\chi = s, {\bf m}|\Psi\right>\right|^2} \ ,
\end{equation}
is the conditional probability of observing the eigenvalue ${\bf m}$ given that $\hat{\chi}$ is observed to have the eigenvalue $s$. The conditional expectation value of the operator $\hat{f}$ given in~(\ref{eq:f-commutes-chi}) is defined as
\begin{equation}\notag
    \E_{\Psi}[f|\chi = s] :=\sum_{\bf m} f(s,{\bf m})\ p_{\Psi}({\bf m}|\chi = s) \ ,
\end{equation}
which, due to~(\ref{eq:f-commutes-chi}), can be written as
\begin{equation}\label{eq:CP-exp}
\E_{\Psi}[f|\chi = s] = \frac{\left<\Psi\left|\hat{f}\hat{P}_{\chi = s}\right|\Psi\right>}{\left<\Psi\left|\hat{P}_{\chi = s}\right|\Psi\right>} \ .
\end{equation}
Although we have used the noninvariant operators $\hat{f}$ and $\hat{P}_{\chi = s}$ and the auxiliary inner product $\braket{\cdot|\cdot}$ in~(\ref{eq:CP-exp}), we can evidently also write it in terms of on-shell operators and the induced inner product $(\cdot|\cdot)$ as defined in~(\ref{eq:induced-IP}). Indeed, due to~(\ref{eq:fis-ID}), Eq.~(\ref{eq:CP-exp}) can be written as
\begin{equation}\label{eq:CP-exp-2}
    \E_{\Psi}[f|\chi = s] = \frac{\left(\Psi\left|\hat{P}_{E = 0}\hat{f}\hat{P}_{\chi = s}\hat{P}_{E = 0}\right|\Psi\right)}{\left(\Psi\left|\hat{P}_{E = 0}\hat{P}_{\chi = s}\hat{P}_{E = 0}\right|\Psi\right)} \ .
\end{equation}
From the general formulas~(\ref{eq:general-inv-quantum}) and~(\ref{eq:general-on-shell-PoP}), we can thus rewrite~(\ref{eq:CP-exp-2}) as
\begin{equation}\notag
    \E_{\Psi}[f|\chi = s] = \frac{\left(\Psi\left|\int_{-\infty}^{\infty}\D\tau\ \e^{\frac\I\hbar\tau\hat{C}}\hat{f}\hat{P}_{\chi = s}\e^{-\frac\I\hbar\tau\hat{C}}\right|\Psi\right)}{\left(\Psi\left|\int_{-\infty}^{\infty}\D\tau\ \e^{\frac\I\hbar\tau\hat{C}}\hat{P}_{\chi = s}\e^{-\frac\I\hbar\tau\hat{C}}\right|\Psi\right)} \ ,
\end{equation}
which is to be compared to the classical formulas~(\ref{eq:relObs-general}) and~(\ref{eq:FP-general}).

We can go further and relate the conditional expectation values to the quantum averages of relational observables. The quantum average of an on-shell observable $\hat{\Ob}$ with respect to an on-shell state $\ket{\Psi}$ is defined as
\begin{equation}
\left<\hat{\Ob}\right>_{\Psi}:= \frac{\left(\Psi\left|\hat{\Ob}\right|\Psi\right)}{\left(\Psi|\Psi\right)} \ .
\end{equation}
Furthermore, we define the (improper) projector onto a given $\sigma$ sector either as in~(\ref{eq:projectors-relObs2}) or as
\begin{equation}
\hat{P}^{\sigma}_{E = 0}:= \sum_{\bf m}\ket{\sigma,{\bf m};s}\bra{\sigma,{\bf m};s} \ ,
\end{equation}
where $\ket{\sigma,{\bf m};s}$ are the eigenstates of $\hat{\Ob}_{(I)}[f|\chi = s]$ defined in~(\ref{eq:eigenstate-relObs}). Thus, we take
\begin{equation}
p_{\Psi}(\sigma) = \left<\hat{P}^{\sigma}_{E = 0}\right>_{\Psi}
\end{equation}
to be the probability that the mechanical system is observed to be in the $\sigma$ sector (e.g., the probability that a free relativistic particle has frequency $\sigma$). Due to~(\ref{eq:relObs-quantum-1}), the quantum average of $\hat{\Ob}_{(I)}[f|\chi = s]$ can be written as
\begin{align*}
\left<\hat{\Ob}_{(I)}[f|\chi = s]\right>_{\Psi} = \sum_{\sigma}p_{\Psi}(\sigma)\frac{\left(\Psi\left|\hat{\Omega}_{\chi}^{\sigma}\hat{f}\hat{P}_{\chi = s}\hat{\Omega}_{\chi}^{\sigma}\right|\Psi\right)}{\left(\Psi\left|\hat{\Omega}_{\chi}^{\sigma}\hat{P}_{\chi = s}\hat{\Omega}_{\chi}^{\sigma}\right|\Psi\right)} \ . 
\end{align*}
If we now define
\begin{equation}\label{eq:Delta-12-Psi}
\ket{\Psi_{\sigma}}:= \hat{\Omega}_{\chi}^{\sigma}\bullet\ket{\Psi} \ ,
\end{equation}
we obtain [cf.~(\ref{eq:omega-delta-12})]
\begin{align*}
&\left<\hat{\Ob}_{(I)}[f|\chi = s]\right>_{\Psi}\\
&= \sum_{\sigma}p_{\Psi}(\sigma)\frac{\left(\Psi_{\sigma}\left|\hat{P}_{E=0}\hat{f}\hat{P}_{\chi = s}\hat{P}_{E=0}\right|\Psi_{\sigma}\right)}{\left(\Psi_{\sigma}\left|\hat{P}_{E=0}\hat{P}_{\chi = s}\hat{P}_{E=0}\right|\Psi_{\sigma}\right)}\\
&= \sum_{\sigma}p_{\Psi}(\sigma)\frac{\left<\Psi_{\sigma}\left|\hat{f}\hat{P}_{\chi = s}\right|\Psi_{\sigma}\right>}{\left<\Psi_{\sigma}\left|\hat{P}_{\chi = s}\right|\Psi_{\sigma}\right>} \ .
\end{align*}
Therefore, we conclude from~(\ref{eq:CP-exp}) that
\begin{equation}\label{eq:CP-relObs}
\left<\hat{\Ob}_{(I)}[f|\chi = s]\right>_{\Psi} = \sum_{\sigma}p_{\Psi}(\sigma)\E_{\Psi_{\sigma}}[f|\chi = s] \ ,
\end{equation}
i.e., the quantum average of the relational observable $\hat{\Ob}_{(I)}[f|\chi = s]$ is the weighted sum of the $\sigma$-sector (``single-frequency'') expectation values of $\hat{f}$ conditioned on $\chi = s$, provided one suitably redefines the states as in~(\ref{eq:Delta-12-Psi}). The weights are the probabilities that the system has a given value of $\sigma$. In particular, if the state $\ket{\Psi}$ has a definite value $\sigma = \sigma_0$, then $p_{\Psi}(\sigma) = \delta_{\sigma,\sigma_0}$ and the quantum average of $\hat{\Ob}_{(I)}[f|\chi = s]$ coincides with a conditional expectation value. In this case, if $\ket{\Psi}$ is normalized,
\begin{equation}\label{eq:conditional-Heisenberg-pic-0}
1 = (\Psi|\Psi) = \left(\Psi\left|\hat{P}^{\sigma_0}_{E = 0}\right|\Psi\right) = 2\pi\hbar\left<\Psi_{\sigma_0}\left|\hat{P}_{\chi = s}\right|\Psi_{\sigma_0}\right> \ , 
\end{equation}
then its overlap with the eigenstates of the relational observable coincides with a conditional probability amplitude; i.e.,
\begin{equation}\label{eq:conditional-Heisenberg-pic}
\begin{aligned}
|(\sigma_0,{\bf m};s|\Psi)|^2 =& 2\pi\hbar|\braket{\chi = s,{\bf m}|\Psi_{\sigma_0}}|^2\\
 =& p_{\Psi_{\sigma_0}}({\bf m}|\chi = s) \ ,
\end{aligned}
\end{equation}
where we used~(\ref{eq:CP}) and~(\ref{eq:conditional-Heisenberg-pic-0}).

This result clarifies the physical interpretation of the quantum relational observable $\hat{\Ob}_{(I)}[f|\chi = s]$. As expected, it is the operator which represents the quantity $\hat{f}$ given the condition that the quantity $\hat{\chi}$ is observed to have the value $s$ (in definite $\sigma$ sectors). A similar result for $\hat{\Ob}_{(II)}[f|\chi = s]$ does not seem to be readily available. Since we regard the use of conditional probabilities intuitively clear and we wish to work with relational observables that have a straightforward physical interpretation, we will thus favor the use of $\hat{\Ob}_{(I)}[f|\chi = s]$ in the remainder of this article.

What is the connection of this result to quantum reference frames? As discussed in Secs.~\ref{sec:QRef} and~\ref{sec:switch}, we interpret the quantum relational observables as invariant descriptions of the dynamics in the reference frame defined by $\hat{\chi}$, i.e., the frame in which the dynamics is described relative to the observed value of $\hat{\chi}$. Likewise, the overlap $(\sigma_0,{\bf m};s|\Psi)$ is interpreted as the representation of the on-shell state in that frame. For definite $\sigma$ sectors, the equivalence of averages of relational observables with conditional expectation values [cf.~(\ref{eq:CP-relObs})] and of the overlap $(\sigma_0,{\bf m};s|\Psi)$ with a conditional probability amplitude [cf.~(\ref{eq:conditional-Heisenberg-pic})] means that these relational quantities describe the quantum dynamics conditioned on the observed value of one of the dynamical fields $\hat{\chi}$. We believe that this corroborates our interpretation of relational observables and their eigenstates in a straightforward, intuitive way. As was argued in Secs.~\ref{sec:QRef} and~\ref{sec:switch}, one can describe the change in reference frames as a change of basis in the physical Hilbert space.\footnote{It is also worthwhile to note that the early formalism presented by Barvinsky in~\cite{Barvinsky} is claimed to be related to the idea of conditional probabilities in quantum cosmology (see page 294 of~\cite{Barvinsky}), but that formalism is restricted to a perturbative expansion in $\hbar$. This restriction is, in principle, not necessary here. We also note that unitarity in~\cite{Barvinsky} was achieved by restricting the theory to a definite $\sigma$ sector. In our formalism, this is also the case in the sense that the conditional expectation values are only equivalent to the averages of relational observables [which evolve unitarily due to~(\ref{eq:gf-Heisenberg}) or~(\ref{eq:gf-prop-unitary})] in definite $\sigma$ sectors [cf.~(\ref{eq:CP-relObs})].}

In fact, Eq.~(\ref{eq:CP-relObs}) shows that there are two equivalent descriptions of the quantum dynamics. One is based on the definite-$\sigma$ conditional expectation values of tensor fields (gauge-fixed point of view), whereas the other is based on the manifestly invariant average of on-shell relational observables in the induced inner product (invariant point of view). This equivalence was first noted in~\cite{Hoehn:Trinity} for a special case, which we analyze in Sec.~\ref{sec:PW}.

To the best of our knowledge, the formalism here presented is new. Although the there is a vast literature on the use of conditional probabilities (see~\cite{PW1,PW2,PW3,PW4,PW5,PW6,PW7,Dolby:2004,Hoehn:Trinity} and references therein) and on the construction of relational observables (see~\cite{Rovelli:1990-1,Rovelli:1990-2,Rovelli:1991,Woodard:1985,HT:book,Woodard:1993,Dittrich:2004,Dittrich:2005,Hoehn:2018-1,Hoehn:2018-2,Chataig:2019,Hoehn:Trinity} for previous proposals and further references) for time-reparametrization invariant systems, the precise connection between the two has so far remained unclear (see, however, the results of~\cite{Hoehn:Trinity}). We hope that the above construction can help bridge the gap between the two approaches and clarify the meaning of quantum relational observables.

What is the significance of this formalism for quantum gravity? There, the construction of well-defined relational observables is notoriously difficult. The present discussion, which is focused on a general mechanical model, cannot be directly applied to field theory, since in the field-theoretic case one must first solve the issue of regularization of the constraint operators and ascertain whether the constraint algebra is anomalous. Nevertheless, the formalism we have presented can be applied to symmetry-reduced models which are frequently used in quantum cosmology. This makes it directly useful to the analysis of quantum-gravitational effects in the early Universe in particular (see the article~\cite{ChataigKraemer}) and of toy-models of quantum gravity in general. Most importantly, we see from~(\ref{eq:CP-relObs}) that it is, in fact, unnecessary to construct relational observables, as long as one is content with computing definite-$\sigma$ conditional expectation values (gauge-fixed point of view).

\subsubsection{\label{sec:invext}Invariant extensions of states}
We have seen that, in the gauge-fixed point of view, one deals with conditional probabilities defined from the on-shell states. There is, of course, an inherent ambiguity in the definition of conditional probabilities that can be seen if one performs the factorization
\begin{equation}\label{eq:CP-fact}
    \braket{\chi = s,{\bf m}|\Psi} = \xi(s)\psi(s,{\bf m}) \ ,
\end{equation}
proposed in~\cite{Hunter:1975}. Here, $\psi(s,{\bf m})$ is called the conditional wave function or the conditional probability amplitude. The conditional probability~(\ref{eq:CP}) only depends on the conditional wave function,
\begin{equation}\label{eq:CP-cond-wf}
    p_{\Psi}({\bf m}|\chi = s) = \frac{\left|\psi(s,{\bf m})\right|^2}{\sum_{\bf m}\left|\psi(s,{\bf m})\right|^2} \ .
\end{equation}
Both the factorization~(\ref{eq:CP-fact}) and the conditional probability~(\ref{eq:CP-cond-wf}) are invariant under the transformations
\begin{equation}
\begin{aligned}
    &\xi(s)\mapsto\e^{\alpha(s)+\I\beta(s)}\xi(s) \ , \\
    &\psi(s,{\bf m})\mapsto\e^{-\alpha(s)-\I\beta(s)}\psi(s,{\bf m}) \ ,
\end{aligned}
\end{equation}
where $\alpha(s),\beta(s)$ are real functions of $s$. Thus, as far as the conditional probabilities are concerned, we are free to choose $\xi(s)$ in a convenient manner. A simple choice is $\xi(s)\equiv1$. If we perform the factorization~(\ref{eq:CP-fact}) for all values of $s$ and a general choice of $\xi(s)$, then the conditional wave function is a solution to the modified constraint equation
\begin{equation}\label{eq:Cxi}
\hat{C}_{\xi}\psi(s,{\bf m}) = 0 \ ,
\end{equation}
where
\begin{equation}
\hat{C}_{\xi}:=\xi^{-1}(s)\hat{C}\left(s,-\I\hbar\frac{\partial}{\partial s},{\bf m},-\I\hbar\frac{\partial}{\partial{\bf m}}\right)\xi(s) \ .
\end{equation}
Now suppose that we know what the conditional probability distribution is at a certain moment of time $s = s_0$ (i.e., for a certain observed value of the field $\hat{\chi}$). Let us call this the ``relative initial data''. Can we determine from this data what is the corresponding on-shell state? Can we subsequently evolve this state with respect to the gauge-fixed time parameter $s$?
 
 The answer to both questions is yes. To see this, suppose that $\ket{\psi}$ is an auxiliary (off-shell) state, such that $\psi(s_0,{\bf m}) = \braket{\chi = s_0, {\bf m}|\psi}$ is a conditional wave function compatible with the relative initial data. Then we can define the on-shell state
\begin{equation}\label{eq:invext}
\ket{\Psi_{\sigma}}:=2\pi\hbar\hat{\Omega}_{\chi}^{\sigma}\bullet\hat{\Omega}_{\chi}^{\sigma}\hat{P}_{\chi = s_0}\ket{\psi} \ .
\end{equation}
Due to~(\ref{eq:conditions-omega-2}), we have the identity
\begin{equation}\label{eq:Pchi-Om2-Pchi}
2\pi\hbar\hat{P}_{\chi = s_0}\hat{\Omega}_{\chi}^{\sigma}\bullet\hat{\Omega}_{\chi}^{\sigma}\hat{P}_{\chi = s_0} = \hat{P}_{\chi = s_0} \ ,
\end{equation}
which implies that
\begin{equation}
\hat{P}_{\chi = s_0}\ket{\Psi_{\sigma}} = \hat{P}_{\chi = s_0}\ket{\psi} \ ,
\end{equation}
or, equivalently,
\begin{equation}
\braket{\chi = s_0,{\bf m}|\Psi_{\sigma}} = \psi(s_0,{\bf m}) \ ,
\end{equation}
i.e., $\ket{\Psi_{\sigma}}$ is an on-shell state that reduces to the initial conditional wave function when projected onto the initial gauge condition.\footnote{The use of gauge conditions in the determination of the initial data for solutions of the quantum constraint equation was also considered in the formalism of~\cite{Barvinsky}.} We refer to such states as invariant extensions. The ``invariantization map'' is a projection in the sense that its square is itself. Indeed, due to~(\ref{eq:Pchi-Om2-Pchi}), we find
\begin{equation}
\begin{aligned}
&(2\pi\hbar)^2\hat{\Omega}_{\chi}^{\sigma}\bullet\hat{\Omega}_{\chi}^{\sigma}\hat{P}_{\chi = s_0}\hat{\Omega}_{\chi}^{\sigma}\bullet\hat{\Omega}_{\chi}^{\sigma}\hat{P}_{\chi = s_0}\\
&=2\pi\hbar\hat{\Omega}_{\chi}^{\sigma}\bullet\hat{\Omega}_{\chi}^{\sigma}\hat{P}_{\chi = s_0} \ .
\end{aligned}
\end{equation}
As the relational observables, the invariant extensions of states can be interpreted in a relational way: they correspond to the value of a given conditional probability amplitude when the field $\hat{\chi}$ is observed to have a certain value. In other words, they encode the relative initial data in a diffeomorphism-invariant fashion. The use of an ``invariantization'' procedure to obtain solutions to the quantum constraint equation was advocated by Woodard in~\cite{Woodard:1993} and, more recently, similar proposals were made in the quantum foundations and and quantum information literature (see~\cite{QRef1,QRef3,QRef4,QRef5} and references therein; we return to this point in Sec.~\ref{sec:PW}).

If the state $\ket{\Psi_{\sigma}}$ is projected onto the gauge condition $\ket{\chi = s,{\bf m}}$ for $s\neq s_0$, we obtain an evolved conditional probability amplitude via the gauge-fixed propagator [cf.~Sec.~\ref{sec:gf-propagator}]. Indeed, due to~(\ref{eq:eigenstate-relObs}) and~(\ref{eq:invext}), we obtain
\begin{equation}\label{eq:invext-propagator}
\begin{aligned}
&\braket{\chi = s,{\bf m}|\Psi_{\sigma}}\\
&= \sum_{{\bf m}_0}(\sigma,{\bf m};s|\sigma,{\bf m}_0;s_0)\psi(s_0,{\bf m}_0)\ .
\end{aligned}
\end{equation}
We will provide an example of an invariant extension and its relational interpretation in Sec.~\ref{sec:flrw}.

\subsubsection{\label{sec:Other-CP}Remark on notation and terminology}
Before we continue, it is worth making a brief remark concerning the notation and terminology used here for conditional probabilities. We note that in~\cite{Dolby:2004}, the probability given in~(\ref{eq:CP}) was denoted by $p_{\Psi}({\bf m}\text{ when }\chi=s)$, and the term ``conditional probability'' was reserved for the different object
\begin{equation}\label{eq:CP-Dolby}
    p_{\Psi}^{\text{alt}}({\bf m}|\chi = s):=\frac{\left<\Psi\left|\hat{P}_{\chi = s}\hat{P}_{E = 0}\hat{P}_{\bf m}\hat{P}_{E = 0}\hat{P}_{\chi = s}\right|\Psi\right>}{\left<\Psi\left|\hat{P}_{\chi=s}\hat{P}_{E = 0}\hat{P}_{\chi = s}\right|\Psi\right>} \ .
\end{equation}
In~\cite{Dolby:2004}, the main motivation to consider~(\ref{eq:CP-Dolby}) was that it leads to a derivation of the usual Schr\"{o}dinger propagator for constraints of the form~(\ref{eq:PW-constraint}). Indeed, if one chooses the gauge condition $\hat{\chi}$ to be canonically conjugate to $\hat{C}_{(1)}$, then one can show that $p_{\Psi}^{\text{alt}}({\bf m}_2\text{ when }\chi = s_2| {\bf m}_1\text{ when }\chi = s_1)$ is the correct transition probability associated with the Schr\"{o}dinger propagator (see~\cite{Dolby:2004} and also the alternative discussion in~\cite{Hoehn:Trinity} for details). In the present article, we are content with defining the conditional probability to be~(\ref{eq:CP}), because we will show that the correct Schr\"{o}dinger propagator is obtained from the gauge-fixed propagator in Sec.~\ref{sec:PW}, without the need to consider~(\ref{eq:CP-Dolby}).

\subsubsection{\label{sec:PW}Relation to the Page-Wootters formalism}
It is now important to note what is the relation of the above construction to the Page-Wootters formalism~\cite{PW1,PW2,PW3,PW4,PW5,PW6,PW7,Dolby:2004,Hoehn:Trinity}, which is the most widely used framework for conditional probabilities in time-reparametrization invariant quantum systems. The goal of the Page-Wootters formalism is to recover the usual notion of evolution with respect to an external time parameter and the time-dependent Schr\"{o}dinger equation from a stationary constraint equation. This can be seen as a particular case of the formalism presented here. 

It is also worthwhile to mention another approach that aims at recovering the usual Schr\"{o}dinger equation, which is the Born-Oppenheimer approach to quantum gravity~\cite{Kiefer:1993}. In this approach, one performs the factorization of the on-shell states as in~(\ref{eq:CP-fact}) and finds that $\xi(s)$ has a Wentzel–Kramers–Brillouin (WKB)-like form, which arises from a weak-coupling expansion of the quantum constraint equation $\hat{C}\ket{\Psi} = 0$. This expansion concerns a weak-coupling between ``heavy'' degrees of freedom (e.g., a laboratory that defines the reference frame) and ``light'' fields (e.g. a subsystem of the model universe). The time-dependent Schr\"{o}dinger equation for the conditional wave function is then recovered from~(\ref{eq:Cxi}) at the lowest order of the expansion. In quantum cosmology, the heavy degrees of freedom are those of the gravitational field, whereas the light fields are those of the matter sector. In this case, the weak-coupling expansion is a (formal) expansion in inverse powers of the Planck mass. We will not pursue this approach here. The reader is referred to~\cite{Chataig:2019-0} for a review and further details in the context of quantum cosmology and to~\cite{ChataigKraemer} for an application to the computation of corrections to the dynamics of primordial fluctuations.

Alternatively, in the Page-Wootters formalism, one often assumes that the laboratory and the system to be studied do not interact. In this case, the constraint is assumed to be of the form~(\ref{eq:PW-constraint}) and we set $\xi(s)\equiv1$. In~(\ref{eq:PW-constraint}), $\hat{C}_{(1)}$ can be interpreted as the laboratory Hamiltonian, whereas $\hat{C}_{(>1)}$ is the Hamiltonian of the system. In the language of the present article, the first step to recover the time-dependent Schr\"{o}dinger equation for the system is to choose a gauge condition $\hat{\chi}$.

It is reasonable to choose $\hat{\chi}$ to be canonically conjugate to $\hat{C}_{(1)}$ (i.e., $\hat{\chi}$ is the ``proper time'' of the laboratory). We assume for simplicity that the spectra of $\hat{C}$, $\hat{C}_{(1)}$ and $\hat{C}_{(>1)}$ are continuous. In this way, if $\ket{E_{(1)},q^i}$ ($i>1$) is an eigenstate of $\hat{C}_{(1)}$, one may define the eigenstates of $\hat{\chi}$ as\footnote{Classically, we assume that the system Hamiltonian $C_{(>1)}$ is positive definite, whereas $C_{(1)}$ is not, such that the constraint~(\ref{eq:PW-constraint}), $C = 0$, is satisfied. In the quantum case, we thus assume that $\hat{C}_{(>1)}$ is positive-definite, whereas the spectrum of $\hat{C}_{(1)}$ ranges over $\mathbb{R}$. The reader is referred to~\cite{Hoehn:Trinity} for a careful discussion of these issues and possible generalizations. In particular, the general case of $\hat{\chi}$ being symmetric but not self-adjoint, such that its associated gauge-fixed time function is a ``covariant positive-operator valued measure (POVM)'', is discussed in~\cite{Hoehn:Trinity}.}
\begin{equation}\label{eq:PW-chi-eigen1}
    \ket{\chi, q^i} = \frac{1}{\sqrt{2\pi\hbar}}\int_{\mathbb{R}}\D E_{(1)}\ \e^{-\frac\I\hbar E_{(1)}\chi}\ket{E_{(1)},q^i} \ ,
\end{equation}
from which one determines
\begin{equation}\label{eq:pre-PW}
    \I\hbar\frac{\D}{\D s}\ket{\chi = s, q^i} =\hat{C}_{(1)} \ket{\chi = s, q^i} \ .
\end{equation}
Given an on-shell state $\ket{\Psi}$, which is a solution to $\hat{C}\ket{\Psi} = 0$, we define the conditional wave function as $\psi(s,q^i):=\braket{\chi = s,q^i|\Psi}$ [cf.~(\ref{eq:CP-fact})]. Using~(\ref{eq:PW-constraint}) and~(\ref{eq:pre-PW}), we find
\begin{equation}\label{eq:PW-Schrodinger}
\begin{aligned}
\I\hbar\frac{\D}{\D s}\psi(s,q^i) &= \left<\chi = s,q^i\left|-\hat{C}_{(1)}\right|\Psi\right>\\
&= \left<\chi = s,q^i\left|\hat{C}_{(>1)}\right|\Psi\right>\\
&=\hat{C}_{(>1)}\left(q,\frac\hbar\I\frac{\partial}{\partial q}\right)\psi(s,q^i) \ .
\end{aligned}
\end{equation}
This is the usual derivation of the Schr\"{o}dinger equation for the system in the Page-Wootters formalism. The dynamics is described with respect to the laboratory proper time $s$.

What about the relational observables? To construct them as in~(\ref{eq:CP-relObs}), we first note that $\hat{C}_{(1)}$ and $\hat{C}_{(>1)}$ form a complete set of commuting invariants. We may thus use the results of Sec.~\ref{sec:relObs-II}. Let $\ket{E_{(1)},E_{(>1)},\bf n}$ be a system of simultaneous eigenstates of $\hat{C}_{(1)}$ and $\hat{C}_{(>1)}$ which are orthornormal with respect to the auxiliary inner product $\braket{\cdot|\cdot}$. We obtain~[cf.~(\ref{eq:function-eigenvalues-Ch})]
\begin{align*}
\hat{C}\ket{E_{(1)},E_{(>1)},\bf n} =&C(E_{(1)},E_{(>1)})\ket{E_{(1)},E_{(>1)},\bf n}\\
=& (E_{(1)}+E_{(>1)})\ket{E_{(1)},E_{(>1)},\bf n} \ ,
\end{align*}
and, therefore, Eq.~(\ref{eq:ChnE}) becomes
\begin{equation}\notag
E_{(1)}+E_{(>1)} = E \ .
\end{equation}
For this simple case, there is no multiplicity in the solution,
\begin{equation}\notag
E_{(1)} = E-E_{(>1)} \ ;
\end{equation}
i.e., there is only one $\sigma$ sector, $\sigma\equiv1$. From this, it also follows that the on-shell states [cf.~(\ref{eq:on-shell-Esigman})]
\begin{equation}\notag
\ket{\epsilon,\bf n}:=\ket{E_{(1)} = -\epsilon,E_{(>1)} = \epsilon,\bf n}
\end{equation}
are orthonormal in the induced inner product [cf.~(\ref{eq:induced-overlap-relObs2})],
\begin{equation}\notag
(\epsilon',{\bf n'}|\epsilon,{\bf n}) = \delta(\epsilon'-\epsilon)\delta({\bf n'-n}) \ .
\end{equation}
The improper projector onto the physical Hilbert space is, thus,
\begin{equation}\label{eq:PW-phys-projector}
\hat{P}_{E = 0}:=\int\D\epsilon\D{\bf n}\ket{\epsilon,\bf n}\bra{\epsilon,\bf n} \ .
\end{equation}
To compute the on-shell Faddeev-Popov operator as in~(\ref{eq:FP-general-quantum}), we note from~(\ref{eq:PW-chi-eigen1}) that
\begin{equation}\label{eq:PW-overlap-chi-0}
    \braket{E_{(1)},E_{(>1)},{\bf n}|\chi,q^i} = \frac{\e^{-\frac\I\hbar E_{(1)}\chi}}{\sqrt{2\pi\hbar}}\braket{E_{(>1)},{\bf n}|q^i} \ ,
\end{equation}
which implies that 
\begin{equation}\label{eq:PW-overlap-chi}
    \begin{aligned}
        &\left<\epsilon',{\bf n}'\left|\hat{P}_{\chi = s}\right|\epsilon,{\bf n}\right>\\
        &= \frac{\e^{\frac\I\hbar\left(\epsilon'-\epsilon\right)s}}{2\pi\hbar}\delta(\epsilon-\epsilon')\delta({\bf n-n'}) \ ,
    \end{aligned}
\end{equation}
if one uses the completeness relation for the $\hat{q}^i$ eigenstates. Using~(\ref{eq:PW-phys-projector}) and~(\ref{eq:PW-overlap-chi}) in the definition~(\ref{eq:FP-general-quantum}), we obtain the result $\hat{\Omega}_{\chi}\equiv\hat{P}_{E = 0}$; i.e., the Faddeev-Popov operator is the identity in the physical Hilbert space. In this case, we note that both choices of factor ordering for the relational observables~(\ref{eq:relObs-quantum-1}) and~(\ref{eq:relObs-quantum-II}) coincide for operators $\hat{f}$ that commute with $\hat{\chi}$, i.e., $\hat{\Ob}_{(I)}[f|\chi = s]=\hat{\Ob}_{(II)}[f|\chi = s]$. Moreover, in this case~(\ref{eq:CP-relObs}) simplifies to
\begin{equation}\label{eq:CP-relObs-PW}
\left<\hat{\Ob}_{(I)}[f|\chi = s]\right>_{\Psi} = \E_{\Psi}[f|\chi = s] \ ;
\end{equation}
i.e., the quantum average of relational observables are exactly the conditional expectation values. This was the case analyzed in~\cite{Hoehn:Trinity}, where the equivalence of the construction of relational observables and the use of conditional probabilities in the Page-Wootters formalism was established for a constraint of the form~(\ref{eq:PW-constraint}) (it is worthwhile to note that, in~\cite{Hoehn:Trinity}, one does not restrict the analysis to the case in which the spectrum of $\hat{C}_{(1)}$ is unbounded in both directions, as we did here for simplicity). The formalism we have presented in Secs.~\ref{sec:relObs-I} and~\ref{sec:relObs-II} can thus be seen as a generalization of this result to more general constraint operators and also to gauge conditions $\hat{\chi}$ that are admissible according to the criterion of Sec.~\ref{sec:relObs-I}, but which are not necessarily the canonical conjugates of the constraint operator.\footnote{It is also worth mentioning that, in~\cite{Gambini:2009}, the use of relational observables and the Page-Wootters formalism were combined. However, the method of~\cite{Gambini:2009} is rather different from the one we present here. In~\cite{Gambini:2009}, one first defines the relational observables, e.g., based on the classical solutions to the field equations, and then one computes conditional probabilities associated with these observables. To achieve this, one integrates over the gauge-fixed time parameter $s$, considered to be unobservable. We do not follow this approach because $s$ is the ``reading of a clock'' (as was also remarked in~\cite{Hoehn:Trinity}); i.e., it is the value of the field $\hat{\chi}$ conditioned on which observations of the other fields are made. Thus, one should not integrate over $s$. Moreover, the construction of the quantum observables here described (see also our previous article~\cite{Chataig:2019}) does not require one to first solve the classical field equations and subsequently quantize the invariant observables, which can be defined directly in the quantum theory, if needed. This is a potential technical advantage over the formalism of~\cite{Gambini:2009}.}

The observables also obey the gauge-fixed Heisenberg equations of motion~(\ref{eq:gf-Heisenberg}), with the invariant Hamiltonian $\hat{H}_{\chi} = -\hat{C}_{(1)}$. In particular, since $\hat{H}_{\chi}\ket{\epsilon,{\bf n}} = \hat{C}_{(>1)}\ket{\epsilon,{\bf n}}$, we obtain
\begin{equation}\notag
\frac{\D}{\D s}\hat{\Ob}_{(I)}[q^i|\chi = s] = \frac{1}{\I\hbar}[\hat{\Ob}_{(I)}[q^i|\chi = s],\hat{C}_{(>1)}] \ (i>1) \ ,
\end{equation}
which are just the usual Heisenberg equations in nonrelativistic quantum mechanics. Furthermore, due to~(\ref{eq:PW-constraint}) and~(\ref{eq:PW-chi-eigen1}), we can also write~(\ref{eq:relObs-quantum-3}) as
\begin{equation}\label{eq:PW-relativization}
\hat{\Ob}_{(I)}[q^i|\chi = s] = \int_{-\infty}^{\infty}\D\tau\ \hat{q}^i(\tau)\otimes\hat{P}_{\chi = s-\tau}\ \ ,
\end{equation}
where
\begin{equation}\notag
\hat{q}^i(\tau) := \e^{\frac{\I}{\hbar}\tau\hat{C}_{(>1)}}\hat{q}^i\e^{-\frac{\I}{\hbar}\tau\hat{C}_{(>1)}} \ .
\end{equation}
Equation~(\ref{eq:PW-relativization}) is reminiscent of the relativization map defined in~\cite{QRef3,QRef4,QRef5} and it is, in fact, the result of the $G$-twirl operation used in~\cite{Hoehn:Trinity} in the context of time-reparametrization invariant quantum mechanics (the $G$-twirl operation has also been used in the context of spatial reference frames~\cite{QRef1}).

Finally, let us mention how, instead of using~(\ref{eq:CP-Dolby}) (as in~\cite{Dolby:2004}, see also the alternative discussion in~\cite{Hoehn:Trinity}), one can recover the usual Schr\"{o}dinger propagator from the gauge-fixed propagator for the Page-Wootters case. From~(\ref{eq:eigenstate-relObs}) with $\hat{\Omega}_{\chi}\equiv\hat{P}_{E = 0}$, we see that the quantum relational observable $\hat{\Ob}[q^i|\chi = s]$ ($i>1$) has the eigenstates
\begin{equation}\label{eq:PW-relational-eigenstates}
    \ket{q^i;s}:=\sqrt{2\pi\hbar}\hat{P}_{E = 0}\ket{\chi = s,q^i} \ .
\end{equation}
Using~(\ref{eq:PW-phys-projector}) and~(\ref{eq:PW-overlap-chi-0}), we obtain the gauge-fixed propagator
\begin{equation}\label{eq:PW-Schrodinger-propagator}
\begin{aligned}
    \left(q'^i;s'|q^j;s\right) &= 2\pi\hbar\left<\chi = s',q'^i\left|\hat{P}_{E = 0}\right|\chi = s,q^j\right>\\
    &=\int\D E_{(>1)}\D{\bf n}\ \braket{q'^i|E_{(>1)},\bf n}\e^{-\frac\I\hbar E_{(>1)}(s'-s)}\\
    &\times\braket{E_{(>1)},{\bf n}|q^j}\\
    &= \left<q'^i\left|\e^{-\frac\I\hbar\hat{C}_{(>1)}(s'-s)}\right|q^j\right> \ ,
\end{aligned}
\end{equation}
where we assumed that $\ket{E_{(>1)},\bf n}$ is a complete orthonormal system in the subspace of the auxiliary Hilbert space spanned by $\ket{q^i}$. Here, $i, j >1$. Equation~(\ref{eq:PW-Schrodinger-propagator}) is the usual Schr\"{o}dinger propagator. It is important to note that, in the formalism here described, the evolution of conditional wave functions, understood as invariant extensions of relative initial data [cf.~Sec.~\ref{sec:invext}], is dictated by the gauge-fixed propagator according to~(\ref{eq:invext-propagator}). The fact that the gauge-fixed propagator reduces to the usual Schr\"{o}dinger propagator in this case is consistent with the fact that the conditional wave function evolves according to the Schr\"{o}dinger equation~(\ref{eq:PW-Schrodinger}).

Thus, we see that the theory of relational observables and the corresponding conditional probabilities presented in Secs.~\ref{sec:relObs-I},~\ref{sec:relObs-II} and~\ref{sec:CP-general} reproduces both the Page-Wootters formalism and the correct Schr\"{o}dinger propagator, without the need to use the alternative definition~(\ref{eq:CP-Dolby}).

In the next section, we will examine a cosmological example.

\section{\label{sec:flrw}FLRW Model}
We are now in a position to apply the general framework developed above to a useful example in cosmology. We consider a closed FLRW model with a massless, minimally coupled and homogeneous scalar field. This model was analyzed before in~\cite{Kiefer:1988}, whereas~\cite{Marolf:1995} dealt with a general analysis of quantum observables and recollapsing universes. In~\cite{Kiefer:1988}, wave packets of on-shell states were constructed, but the precise definition of the physical Hilbert space and the quantum observables was not given. In~\cite{Marolf:1995}, a general analysis of the quantum observables and the induced inner product was carried out, but no connection to conditional probabilities was established. In fact, as was remarked in~\cite{Chataig:2019}, the quantum observables were defined in~\cite{Marolf:1995} in such a way that the Faddeev-Popov resolution of the identity $ \Ob[1|\chi = s] = 1$ [cf.~(\ref{eq:FP-general-0})] was not obtained. We consider this to be undesirable. Here, on the contrary, we take this resolution of the identity to be one of the defining properties of the formalism we have developed, both in its original version presented in~\cite{Chataig:2019} and in its revised version presented in Secs.~\ref{sec:relObs-I} and~\ref{sec:relObs-II}. Thus, our analysis differs from the previous ones in terms of the precise definition of the quantum observables, and we establish their relation to conditional probabilities.

\subsection{Classical theory}
The action in a spacetime region $\mathcal{M}$ is
\begin{align}
    S &= S_{\mathcal{M}}+S_{\partial\mathcal{M}} \ , \label{eq:total-action}\\
    S_{\mathcal{M}} &= \int_{\mathcal{M}}\D^4x\sqrt{-g}\left[\frac{1}{2\kappa}R-\frac{1}{2}\left(\nabla\phi\right)^2\right] \ , \\
    S_{\partial\mathcal{M}} &= -\frac{1}{\kappa}\int_{\partial\mathcal{M}}\D^3x\sqrt{h}\ K \ ,
\end{align}
where $\kappa = \frac{8\pi G}{c^4}$, $R$ is the Ricci scalar and $h, K$ are the determinant of the induced metric and the trace of the extrinsic curvature of the boundary, respectively. We assume the line element
\begin{equation}\label{eq:metric-flrw}
    \D s^2 = -N^2(\tau)\D\tau^2+a^2(\tau)\D\Omega_3^2 \ ,
\end{equation}
where $\D\Omega_3^2 = \D\chi^2+\sin^2\chi\left(\D\theta^2+\sin^2\theta\D\varphi^2\right)$ is the line element on $\mathbb{S}^3$. From~(\ref{eq:metric-flrw}), we find (see, for instance~\cite{Kiefer:book})
\begin{align}
    R &= \frac{6}{N^2}\left[\frac{\ddot{a}}{a}-\frac{\dot{a}\dot{N}}{aN}+\left(\frac{\dot{a}}{a}\right)^2\right]+\frac{6}{a^2} \ , \label{eq:Ricci-flrw}\\
    K &= \frac{3\dot{a}}{aN} \ .
\end{align}
Assuming $N(\tau)>0$, we also have $\sqrt{-g} = N a^3\sin^2\chi\sin\theta$. An integration by parts of the first term in~(\ref{eq:Ricci-flrw}) yields the symmetry-reduced action\footnote{The reason we can impose the symmetry reduction directly at the level of the action and not only in the field equations is that this homogeneous, isotropic model satisfies the {\it symmetric criticality principle} (see~\cite{Kiefer:book,Palais:1979,Torre:1999,Fels:2002} for more details).}
\begin{equation}\label{eq:action-red}
    S = 2\pi^2\int_{\tau_0}^{\tau_1}\D\tau \left(-3\frac{a\dot{a}^2}{\kappa N}+\frac{3Na}{\kappa}+\frac{a^3}{2}\frac{\dot{\phi}^2}{N}\right)\ .
\end{equation}
It is convenient to choose units in which $\frac{6\pi^2}{\kappa} = \frac{1}{2}$ and to make the following redefinitions
\begin{equation}
    \begin{aligned}
        a(\tau) &= \e^{\alpha(\tau)} \ , \\
        N(\tau) &= \e^{3\alpha(\tau)}e(\tau) \ , \\
        \phi(\tau) &\to \frac{1}{\sqrt{2}\pi}\phi(\tau) \ ,
    \end{aligned}
\end{equation}
such that~(\ref{eq:action-red}) becomes
\begin{equation}
    S = \int_{\tau_0}^{\tau_1}\D\tau \left(-\frac{\dot{\alpha}^2}{2 e}+\frac{\dot{\phi}^2}{2e}+\frac{e}{2}\e^{4\alpha}\right)\ .
\end{equation}
After the usual Legendre transform (with $p_e = 0$), we obtain the action in Hamiltonian form
\begin{equation}\label{eq:action-red-2}
    S = \int_{\tau_0}^{\tau_1}\D\tau \left(p_{\alpha}\dot{\alpha}+p_{\phi}\dot{\phi}-e(\tau)C\right)\ ,
\end{equation}
with the constraint
\begin{equation}\label{eq:flrw-C}
    C = -\frac{p_{\alpha}^2}{2}+\frac{p_{\phi}^2}{2}-\frac{\e^{4\alpha}}{2} \ .
\end{equation}
Equation~(\ref{eq:action-red-2}) is of the form~(\ref{eq:action-general}). The symmetry-reduced field equations are
\begin{equation}\label{eq:flrw-eom}
\begin{aligned}
    \dot{\alpha} &=-e(\tau)p_{\alpha} \ , \ \dot{p}_{\alpha} = 2e(\tau)\e^{4\alpha} \ , \\
    \dot{\phi} &= e(\tau)p_{\phi} \ , \ \dot{p}_{\phi} = 0 \ , \\
    0 &= -\frac{p_{\alpha}^2}{2}+\frac{p_{\phi}^2}{2}-\frac{\e^{4\alpha}}{2} \ .
\end{aligned}
\end{equation}
For any choice of time coordinate $\tau$ [for any choice of $e(\tau)$], we can solve the above system of equations in a relational manner, i.e. by describing the dynamics of one field in terms of another. In this way, we can conveniently rewrite~(\ref{eq:flrw-eom}) as follows:
\begin{equation}\label{eq:flrw-eom-relational}
    \begin{aligned}
    \dot{\alpha} &= -\frac{p_{\alpha}}{p_{\phi}}\dot{\phi} \ , \ \dot{p}_{\alpha} = \frac{2\e^{4\alpha}}{p_{\phi}}\dot{\phi} \ , \\
    p_{\phi} &= \sigma\sqrt{p_{\alpha}^2+\e^{4\alpha}} \equiv\sigma|k| \ ,
    \end{aligned}
\end{equation}
where $\sigma = \pm1$ labels the different multiplicity (frequency) sectors and $k$ is a constant of integration. The relational solution of~(\ref{eq:flrw-eom-relational}) is
\begin{align*}
    a^2(\tau) &=\frac{|k|}{\cosh\left[2\sigma(\phi(\tau)-s)+\mathrm{arctanh}\left(\frac{\left.p_{\alpha}\right|_{\phi(\tau) = s}}{|k|}\right)\right]}\ , \\
    p_{\alpha}(\tau) &= |k|\mathrm{tanh}\left[2\sigma(\phi(\tau)-s)+\mathrm{arctanh}\left(\frac{\left.p_{\alpha}\right|_{\phi(\tau) = s}}{|k|}\right)\right] \! ,
\end{align*}
and it is valid for any choice of $\tau$. We note that this solution depends on $|k|$ and $\left.p_{\alpha}\right|_{\phi(\tau) = s}$. From~(\ref{eq:flrw-eom-relational}), we can replace $|k|$ by
\begin{equation}\notag
    |k| = \sqrt{\left.p_{\alpha}^2\right|_{\phi(\tau) = s}+\left.a^4\right|_{\phi(\tau) = s}} \ .
\end{equation}
Therefore, the relational solution only depends on the quantities $\left.\alpha\right|_{\phi(\tau) = s}$ and $\left.p_{\alpha}\right|_{\phi(\tau) = s}$, which label the physical or reduced phase space of the theory. These quantities are the relational observables. We can find explicit expressions for them by inverting the relational solution. For example, we find
\begin{equation}\label{eq:flrw-classical-a2-0}
    \left.a^2\right|_{\phi(\tau)=s} =  \frac{|k|}{\cosh\left[2\sigma(s-\phi(\tau))+\mathrm{arctanh}\left(\frac{p_{\alpha}(\tau)}{|k|}\right)\right]}\ ,
\end{equation}
and similarly for $\left.p_{\alpha}\right|_{\phi(\tau) = s}$. It is straightforward to verify that these observables are invariant under on-shell diffeomorphisms of $\tau$ for a fixed value of $s$. Indeed, we obtain from~(\ref{eq:flrw-eom-relational}) the on-shell identity
\begin{equation}\label{eq:flrw-on-shell-id}
    \frac{\D}{\D\tau}\left[2\sigma(s-\phi(\tau))+\mathrm{arctanh}\left(\frac{p_{\alpha}(\tau)}{|k|}\right)\right] = 0 \ .
\end{equation}
Thus, under an infinitesimal diffeomorphism, $\left.a^2\right|_{\phi(\tau)=s}$ transforms as
\begin{equation}\notag
    \delta_{\epsilon(\tau)} \left.a^2\right|_{\phi(\tau)=s} = \epsilon(\tau)\frac{\D}{\D\tau} \left.a^2\right|_{\phi(\tau)=s} = 0 \ ,
\end{equation}
where we used~(\ref{eq:flrw-on-shell-id}). An analogous calculation shows that $\left.p_{\alpha}\right|_{\phi(\tau) = s}$ is also an invariant. The evolution of the relational observables in terms of the variable $s$ can be expressed in terms of Poisson brackets. For example, we find
\begin{equation}\label{eq:flrw-gf-eom}
\begin{aligned}
    \frac{\D}{\D s}\left.a^2\right|_{\phi(\tau) = s} &= -\frac{\partial}{\partial\phi(\tau)}\left.a^2\right|_{\phi(\tau) = s}\\
    &= \left\{p_{\phi},\left.a^2\right|_{\phi(\tau) = s}\right\} \ .
\end{aligned}
\end{equation}
Equation~(\ref{eq:flrw-gf-eom}) is the gauge-fixed equation of motion for the field $a^2$ (cf.~Sec.~\ref{sec:classical-general} and~\cite{Chataig:2019}). In Sec.~\ref{sec:flrw-quantum}, we will show how the quantum relational observables obey the quantum version of~(\ref{eq:flrw-gf-eom}), which is the gauge-fixed Heisenberg equation of motion [cf.~Sec.~\ref{sec:relObs-II}].

Moreover, the (reduced phase-space) evolution of $\left.a^2\right|_{\phi = s}$ with respect to $s$ can be expressed entirely in terms of the relational observables, without reference to the noninvariant fields $\phi(\tau)$ and $p_{\alpha}(\tau)$. This can be obtained directly from~(\ref{eq:flrw-classical-a2-0}), by evaluating it at different values of $s$. We find
\begin{equation}\label{eq:flrw-classical-a2}
    \left.a^2\right|_{\phi = s}\! =\! \frac{|k|}{\cosh\left[2\sigma(s-s_0)+\mathrm{arccosh}\left(\frac{|k|}{\left.a^2\right|_{\phi = s_0}}\right)\right]} \ .
\end{equation}
This equation shows that this model universe recollapses; i.e., the scale factor expands to a maximum value and starts to contract again.

Finally, as all the variables in this model are worldline scalars, we can express the relational observables as integrals as in~(\ref{eq:relObs-general}) (see~\cite{Chataig:2019} for a generalization to worldline one-forms). We obtain
\begin{equation}
\begin{aligned}
    \Ob[a^2|\phi = s] &:= \left.a^2\right|_{\phi(\tau) = s}\\
    &=\Delta_{\phi}\int_{-\infty}^{\infty}\D\tau\ \delta(\phi(\tau)-s)a^2(\tau) \ ,
\end{aligned}
\end{equation}
where
\begin{equation}
    \begin{aligned}
    \Delta_{\phi}^{-1}:= \int_{-\infty}^{\infty}\D\tau\ \delta(\phi(\tau)-s) \ .
    \end{aligned}
\end{equation}
Therefore, we see that the above relational observables are obtained by choosing the  gauge condition to be [cf.~(\ref{eq:gauge-condition-general})]
\begin{equation}\label{eq:classical-gauge-phi}
    \chi(\alpha(\tau),p_{\alpha}(\tau),\phi(\tau),p_{\phi}(\tau))\equiv\phi(\tau) \ .
\end{equation}
In this way, the level sets of $\phi(\tau)$ define a new time coordinate $s$.

\subsection{\label{sec:flrw-quantum}Quantum theory}
\subsubsection{The physical Hilbert space}
Let us take the auxiliary (off-shell) Hilbert space of the theory to be $L^2(\mathbb{R}^2,\D\alpha\D\phi)$. The quantum constraint is~[cf.~(\ref{eq:flrw-C})]
\begin{equation}\label{eq:flrw-C-quantum}
    \hat{C} := -\frac{\hat{p}_{\alpha}^2}{2}+\frac{\hat{p}_{\phi}^2}{2}-\frac{\e^{4\hat{\alpha}}}{2} \ .
\end{equation}
In order to define the on-shell states and the induced inner product, we consider the eigenvalue problem of~(\ref{eq:flrw-C-quantum}),
\begin{equation}\label{eq:flrw-C-eigen}
    \left(\frac{\hbar^2}{2}\frac{\partial^2}{\partial\alpha^2}-\frac{\hbar^2}{2}\frac{\partial^2}{\partial\phi^2}-\frac{\e^{4\alpha}}{2}\right)\Psi(\alpha,\phi) = E\Psi(\alpha,\phi) \ .
\end{equation}
For $E\geq 0$, we define $E = \frac{\lambda^2}{2}$, and we find the eigenstates
\begin{equation}\label{eq:flrw-eigenstate-1}
\begin{aligned}
    \braket{\alpha,\phi|E,\sigma,k}&:= \Psi_{E,\sigma,k}(\alpha,\phi)\\
    &=\! \exp\left(\frac\I\hbar\sigma\sqrt{k^2+\lambda^2}\phi\right)K_{\frac{\I k}{2\hbar}}\left(\frac{\e^{2\alpha}}{2\hbar}\right)\! ,
\end{aligned}
\end{equation}
where $\sigma = \pm1$ and $K_{\frac{\I k}{2\hbar}}\left(\frac{\e^{2\alpha}}{2\hbar}\right)$ is a modified Bessel function. This solution was chosen to satisfy the boundary condition $\lim_{\alpha\to\infty}\Psi_{E,\sigma,k} = 0$. In what follows, we will make use of the following identities~\cite{Bessel}:
\begin{align}
    &\overline{K_{\I\nu}(x)} = K_{-\I\nu}(x) = K_{\I\nu}(x) \ ,\label{eq:Bessel-1}\\
    &\int_{\mathbb{R}}\D\alpha K_{\I\nu'}\left(\frac{\e^{2\alpha}}{2\hbar}\right)K_{\I\nu}\left(\frac{\e^{2\alpha}}{2\hbar}\right) = \frac{\pi^2\delta(|\nu|-|\nu'|)}{4\nu\sinh(\pi\nu)} ,\label{eq:Bessel-2}\\
    &K_{\I\nu}(x)= \frac{1}{2}\int_{-\infty}^{\infty}\D y\ \e^{-x\cosh y}\cos(\nu y)\ . \label{eq:Bessel-3}
\end{align}
Using~(\ref{eq:Bessel-1}) and~(\ref{eq:Bessel-2}), we obtain the auxiliary inner product
\begin{equation}\label{eq:flrw-aux-P}
\braket{E',\sigma',k'|E,\sigma,k}= \delta\left(E'-E\right)\left(E,\sigma',k'|E,\sigma,k\right) \ ,
\end{equation}
where
\begin{equation}\label{eq:flrw-IP-1}
    \!\!\!\!\left(E,\sigma',k'|E,\sigma,k\right)\!=\! \frac{2\pi^3\hbar^3\sqrt{k^2\!+\!\lambda^2}}{k\sinh\left(\frac{\pi k}{2\hbar}\right)}\delta_{\sigma',\sigma}\delta(|k'|\!-\!|k|) \ .
\end{equation}
One can also repeat this analysis for $E\leq0$ by setting $E = -\frac{\lambda^2}{2}$. In this case, instead of~(\ref{eq:flrw-eigenstate-1}), we obtain
\begin{equation}\label{eq:flrw-eigenstate-2}
    \braket{\alpha,\phi|E,\sigma,k} := \e^{\frac\I\hbar \sigma|k|\phi}K_{\I\nu(\lambda,k)}\left(\frac{\e^{2\alpha}}{2\hbar}\right) \ ,
\end{equation}
where
\begin{equation}\notag
    \nu(\lambda,k):=\frac{1}{2\hbar}\sqrt{k^2+\lambda^2} \ .
\end{equation}
The auxiliary inner product of two of the eigenstates given in~(\ref{eq:flrw-eigenstate-2}) is of the same form as~(\ref{eq:flrw-aux-P}) with
\begin{equation}\label{eq:flrw-IP-2}
    \!\!\!\!\left(E,\sigma',k'|E,\sigma,k\right)\!=\! \frac{2\pi^3\hbar^3}{\sinh\left(\frac{\pi \sqrt{k^2\!+\!\lambda^2}}{2\hbar}\right)}\delta_{\sigma',\sigma}\delta(|k'|\!-\!|k|) \ .
\end{equation}
Finally, we can define the on-shell states
\begin{equation}\label{eq:flrw-on-shell-states}
    \ket{\sigma,k}:=\NN(k)\ket{E = 0, \sigma, k} \ ,
\end{equation}
where $\NN(k)$ is a normalization factor,
\begin{equation}\label{eq:flrw-normalization}
    \NN(k) := \left[\frac{\sinh\left(\frac{\pi|k|}{2\hbar}\right)}{4\pi^3\hbar^3}\right]^{\frac{1}{2}} \ .
\end{equation}
In this way, the induced inner product of the on-shell states~(\ref{eq:flrw-on-shell-states}) can be found by taking the $\lambda\to0$ limit of~(\ref{eq:flrw-IP-1}) or~(\ref{eq:flrw-IP-2}). The result is
\begin{equation}\label{eq:flrw-IP}
    \left(\sigma',k'|\sigma,k\right) = \frac{1}{2}\delta_{\sigma',\sigma}\delta(|k'|-|k|) \ .
\end{equation}
The physical Hilbert space is then defined to be the vector space of superpositions of~(\ref{eq:flrw-on-shell-states}) that are square-integrable with respect to the induced inner product~(\ref{eq:flrw-IP}). The (improper) projector onto the physical Hilbert space is
\begin{equation}\label{eq:flrw-projector}
    \hat{P}_{E = 0}:=\sum_{\sigma = \pm}\int_{-\infty}^{\infty}\D k\ \ket{\sigma,k}\bra{\sigma,k} \ .
\end{equation}

\subsubsection{\label{sec:flrw-relObs-I}Quantum relational observables}
Our goal is now to construct the quantum analogue of~(\ref{eq:flrw-classical-a2-0}) and to show that it obeys a gauge-fixed Heisenberg equation [cf.~(\ref{eq:gf-Heisenberg})] that is the quantum version of~(\ref{eq:flrw-gf-eom}). We thus consider the gauge condition $\hat{\phi}$ [cf.~\ref{eq:classical-gauge-phi}]. Since its momentum is already an invariant, i.e., $\hat{p}_{\phi}$ commutes with the constraint operator, we will use the formalism of Sec.~\ref{sec:relObs-II}.

Let us define the states $\ket{k,p_{\phi}}$ to be the simultaneous orthonormal eigenstates of the complete set of commuting invariants $\hat{p}_{\phi}$ and $\hat{C}_{\alpha} = \frac{\hat{p}_{\phi}^2}{2}-\hat{C}$, where
\begin{align*}
\hat{C}_{\alpha} &:= \frac{\hat{p}_{\alpha}^2}{2}+\frac{\e^{4\alpha}}{2} \ , \\
\hat{C}_{\alpha}\ket{k,p_{\phi}} &= \frac{k^2}{2}\ket{k,p_{\phi}}\ , \\
\braket{\alpha,\phi|k,p_{\phi}} &:= |k|^{\frac{1}{2}}\mathcal{N}(k)\e^{\frac{\I}{\hbar}p_{\phi}\phi}K_{\frac{\I k}{2\hbar}}\left(\frac{\e^{2\alpha}}{2\hbar}\right) \ ,
\end{align*}
and [cf.~(\ref{eq:ChnE})]
\begin{equation}\label{eq:ChnE-flrw}
\hat{C}\ket{k,p_{\phi}} = \left(\frac{p_{\phi}^2}{2}-\frac{k^2}{2}\right)\ket{k,p_{\phi}} \ . 
\end{equation}
We thus find from~(\ref{eq:ChnE-flrw}) the on-shell condition [cf.~(\ref{eq:quantum-freq-sectors})]
\begin{equation}\notag
p_{\phi} = -H_{\phi}^{\sigma} = \sigma|k| \ (\sigma=\pm1) \ ,
\end{equation}
and the on-shell states [cf.~(\ref{eq:on-shell-Esigman}) and~(\ref{eq:induced-overlap-relObs2-normalization})]
\begin{equation}\notag
\ket{\sigma,k} = |k|^{-\frac{1}{2}}\ket{k,p_{\phi}}_{p_{\phi} = \sigma|k|} \ , 
\end{equation}
which are in accordance with~(\ref{eq:flrw-on-shell-states}). From~(\ref{eq:matrix-element-FP}) and~(\ref{eq:FP-general-quantum-powers}), we obtain the Faddeev-Popov operator
\begin{equation}\label{eq:flrw-FP-powers}
\hat{\Omega}_{\phi}^{\sigma}:= \int_{\mathbb{R}}\D k\ |k|^{\frac{1}{2}}\ket{\sigma,k}\bra{\sigma,k} \ .
\end{equation}
We can now construct the observable
\begin{equation}\label{eq:flrw-q-relObs-1}
\hat{\Ob}_{(I)}[f(\alpha)|\phi = s] := \sum_{\sigma=\pm}\int_{\mathbb{R}}\D\alpha f(\alpha)\ket{\sigma,\alpha;s}\bra{\sigma,\alpha;s} \ .
\end{equation}
Its eigenstates read [cf.~(\ref{eq:eigenstate-relObs})]
\begin{equation}\label{eq:flrw-eigenstate-relObs}
\ket{\sigma,\alpha;s}:=\sqrt{2\pi\hbar}\hat{\Omega}_{\phi}^{\sigma}\ket{\alpha,\phi = s} \ .
\end{equation}
To verify that they form a complete system in the physical Hilbert space [cf.~(\ref{eq:conditions-omega-1})], we must calculate the matrix element $\left(\sigma',k'\left|\hat{\Ob}[1|\phi = s]\right|\sigma, k\right)$, which is equal to
\begin{equation}\label{eq:flrw-O1-matrix-element}
    \sum_{\sigma''=\pm}\int_{\mathbb{R}}\D\alpha\left(\sigma',k'|\sigma'',\alpha;s\right)\left(\sigma'',\alpha;s|\sigma, k\right) \ .
\end{equation}
If we insert 
\begin{equation}\label{eq:flrw-relObs-eigenstates-2}
\begin{aligned}
    &\left(\sigma',k'|\sigma, \alpha;s\right)\\
    &= \sqrt{2\pi\hbar}\delta_{\sigma',\sigma}\NN(k')K_{\frac{\I k'}{2\hbar}}\left(\frac{\e^{2\alpha}}{2\hbar}\right)|k'|^{\frac12}\e^{-\frac\I\hbar\sigma'|k'|s} \ ,
\end{aligned}
\end{equation}
into~(\ref{eq:flrw-O1-matrix-element}) and use~(\ref{eq:Bessel-2}) and the definition of the normalization factor given in~(\ref{eq:flrw-normalization}), we find
\begin{equation}\label{eq:flrw-FP}
    \left(\sigma',k'\left|\hat{\Ob}_{(I)}[1|\phi = s]\right|\sigma, k\right) = \frac{1}{2}\delta_{\sigma',\sigma}\delta(|k'|-|k|) \ .
\end{equation}
Thus, we conclude that $\hat{\Ob}_{(I)}[1|\phi = s]$ is the identity in the physical Hilbert space, as it should be [cf.~(\ref{eq:relObs-quantum-1-FP})].

What about the dynamics of the general operator $\hat{\Ob}_{(I)}[f(\alpha)|\phi = s]$? From~(\ref{eq:flrw-relObs-eigenstates-2}), we find
\begin{equation}\label{eq:flrw-relObs-eigen-derivative-1}
    \left(\sigma',k'\left|\I\hbar\frac{\partial}{\partial s}\right|\sigma, \alpha;s\right) = \sigma'|k'|\left(\sigma',k'|\sigma, \alpha;s\right) \ .
\end{equation}
Moreover, the identity
\begin{equation}\label{eq:flrw-on-shell-pphi}
    \hat{p}_{\phi}\ket{\sigma,k} = \sigma|k|\ket{\sigma,k} \ ,
\end{equation}
holds due to the definitions given in~(\ref{eq:flrw-eigenstate-1}) and~(\ref{eq:flrw-on-shell-states}). Thus, we can use~(\ref{eq:flrw-on-shell-pphi}) in~(\ref{eq:flrw-relObs-eigen-derivative-1}) to obtain
\begin{equation}\label{eq:flrw-relObs-eigen-derivative-2}
    \I\hbar\frac{\partial}{\partial s}\ket{\sigma,\alpha;s} = \hat{p}_{\phi}\ket{\sigma,\alpha;s} \ ,
\end{equation}
which, due to~(\ref{eq:flrw-q-relObs-1}), implies that the operator $\hat{\Ob}_{(I)}[f(\alpha)|\phi = s]$ is a solution to the gauge-fixed Heisenberg equation [cf.~(\ref{eq:gf-Heisenberg})]
\begin{equation}\label{eq:flrw-gf-Heisenberg}
    \I\hbar\frac{\partial}{\partial s}\hat{\Ob}_{(I)}[f(\alpha)|\phi = s] = \left[\hat{p}_{\phi},\hat{\Ob}_{(I)}[f(\alpha)|\phi = s]\right] \ ,
\end{equation}
which is the quantum version of~(\ref{eq:flrw-gf-eom}) for the particular case $f(\alpha) = \e^{2\alpha}$. As $\hat{p}_{\phi}$ is self-adjoint with respect to the auxiliary (off-shell) inner product $\braket{\cdot|\cdot}$ and it commutes with the constraint operator, it is self-adjoint with respect to the induced inner product $\left(\cdot|\cdot\right)$. Thus, the dynamics described by~(\ref{eq:flrw-gf-Heisenberg}) is unitary. The physical Hamiltonian is $\hat{H}_{\phi}=-\hat{p}_{\phi}$.

\subsubsection{\label{sec:flrw-rel-dynamics}Relational quantum dynamics}
Let us now examine the relational quantum dynamics of this cosmological model by applying the formalism of the gauge-fixed propagator [cf.~Sec.~\ref{sec:gf-propagator}] and invariant extensions [cf.~Sec.~\ref{sec:invext}] to a simplified example. Suppose the relative initial data (at $\phi = s_0$) is the conditional wave function
\begin{equation}\label{eq:flrw-relative-initial-0}
\braket{\alpha,\phi = s_0|\psi} = \psi(\alpha) = \int_{\mathbb{R}}\D k\ \psi(k)K_{\frac{\I k}{2\hbar}}\left(\frac{\e^{2\alpha}}{2\hbar}\right) \ ,
\end{equation}
where $\psi(k)$ is an even function of $k$, possibly also dependent on $s_0$. Its invariant extension in a given $\sigma$ sector is [cf.~(\ref{eq:invext}) and~(\ref{eq:invext-propagator})]
\begin{equation}\label{eq:flrw-invext-0}
\braket{\alpha,\phi|\Psi_{\sigma}}:=\int_{\mathbb{R}}\D\alpha_0\ (\sigma,\alpha;\phi|\sigma,\alpha_0;s_0)\psi(\alpha_0) \ ,
\end{equation}
where $(\sigma,\alpha;\phi|\sigma,\alpha_0;s_0)$ is the $\sigma$-sector gauge-fixed propagator. For convenience, we define $\ket{\Psi}:=\frac{1}{2}\sum_{\sigma}\ket{\Psi_{\sigma}}$. The gauge-fixed propagator then reads [cf.~(\ref{eq:flrw-FP-powers})]
\begin{equation}\label{eq:flrw-gf-propagator-0}
\begin{aligned}
&\frac{1}{2}\sum_{\sigma}(\sigma,\alpha;\phi|\sigma,\alpha_0;s_0)\\
&:=\pi\hbar\sum_{\sigma}\left<\alpha,\phi\left|\hat{\Omega}_{\phi}^{\sigma}\bullet\hat{\Omega}_{\phi}^{\sigma}\right|\alpha_0,s_0\right>\\
&=\pi\hbar\left<\alpha,\phi\left||\hat{p}_{\phi}|\hat{P}_{E = 0}\right|\alpha_0,s_0\right>\\
&=2\pi\hbar\int_{\mathbb{R}}\D k\ \mathcal{N}^2|k|\cos\left[\frac{k}{\hbar}(\phi-s_0)\right]K_{\frac{\I k}{2\hbar}}(x)K_{\frac{\I k}{2\hbar}}\left(x_0\right)\ ,
\end{aligned}
\end{equation}
where we denoted $x = \frac{\e^{2\alpha}}{2\hbar}$ (similarly for $x_0$) for brevity. Using~(\ref{eq:Bessel-2}),~(\ref{eq:flrw-relative-initial-0}) and~(\ref{eq:flrw-gf-propagator-0}), the invariant extension $\braket{\alpha,\phi|\Psi} \equiv \Psi(\alpha,\phi)$ is found from~(\ref{eq:flrw-invext-0}) to be
\begin{equation}\label{eq:flrw-invext}
\Psi(\alpha,\phi) = \int_{\mathbb{R}}\D k\ \psi(k)\cos\left[\frac{k}{\hbar}(\phi-s_0)\right]K_{\frac{\I k}{2\hbar}}\left(\frac{\e^{2\alpha}}{2\hbar}\right) \ .
\end{equation}
One may verify that~(\ref{eq:flrw-invext}) is a solution to the constraint equation~(\ref{eq:flrw-C-quantum}) that reduces to~(\ref{eq:flrw-relative-initial-0}) if $\phi = s_0$. As an example, let us consider the case
\begin{equation}\label{eq:flrw-invext-example}
\psi(k) = \frac{k}{\hbar}\sin\left(\frac{k}{\hbar}c_0\right) \ ,
\end{equation}
where $c_0$ is a real constant. Using~(\ref{eq:Bessel-3}) and~(\ref{eq:flrw-invext-example}), Eq.~(\ref{eq:flrw-invext}) becomes
\begin{equation}\label{eq:flrw-invext-1}
\begin{aligned}
&\Psi(\alpha,\phi)\\
&=  -\pi\hbar\sum_{\sigma = \pm}\frac{\partial}{\partial c_0}\exp\left\{-\frac{\e^{2\alpha}}{2\hbar}\cosh\left[2\sigma\left(\phi-s_0\right)+2c_0\right]\right\} \ .
\end{aligned}
\end{equation}
This is an invariant extension of the initial conditional wave function (relative initial data)
\begin{equation}\label{eq:flrw-relative-initial-1}
\Psi(\alpha,s_0) =  2\pi\e^{2\alpha}\sinh(2c_0)\exp\left[-\frac{\e^{2\alpha}}{2\hbar}\cosh(2c_0)\right] \ .
\end{equation}
For a general value $\phi = s$, Eq.~(\ref{eq:flrw-invext-1}) corresponds to the superposition of two conditional probability amplitudes, each of which leads to a conditional exponential distribution of the scale factor squared, $a^2 = \e^{2\alpha}$, with the corresponding mean values
\begin{equation}\label{eq:flrw-a2-quantum-mean}
\left.a^2\right|_{\text{mean}} = \frac{2\hbar}{\cosh\left[2\sigma\left(s-s_0\right)+2c_0\right]} \ ,
\end{equation}
which are to be compared to the classical solution~(\ref{eq:flrw-classical-a2}). In particular, Eq.~(\ref{eq:flrw-a2-quantum-mean}) also exhibits a recollapse, i.e., $\lim_{s\to\pm\infty}\left.a^2\right|_{\text{mean}} = 0$. One might then conclude that the singularity is not avoided in the quantum theory here described. However, at least for the state~(\ref{eq:flrw-invext-1}), the conditional probability vanishes in the region of the classical singularity; i.e., it satisfies
\begin{equation}\label{eq:flrw-CP-DeWitt}
\lim_{\alpha\to\pm\infty}p_{\Psi}(\alpha|\phi = s) = 0 \ .
\end{equation}
This is (a probabilitistic version of) DeWitt's criterion for singularity avoidance~\cite{Kiefer:book,DeWitt:1967} and can be interpreted as the statement: ``given that $\phi$ is observed to have the value $s$, the probability that $a^2 = 0$ is zero''.

\section{\label{sec:conclusions}Conclusions}
Despite decades of great effort, there are still some crucial technical and conceptual challenges that face candidate theories of quantum gravity. One of them is the precise understanding of what probabilitistic predictions a diffeormorphism-invariant quantum theory can make. How do we use the wave function(al) of gravitational and matter fields to predict the probabilities of certain observations? Another issue is the proper comprehension of the diffeomorphism symmetry in the quantum realm. What observables act on the physical Hilbert space and what is their physical interpretation?

In this article, we have brought these two topics together by describing a formalism of construction of quantum relational observables, the averages of which can be related to conditional expectation values of tensor fields. The formalism here described can be seen as an extension of certain results of~\cite{Hoehn:Trinity,Barvinsky,QRef3,QRef4}, although adapted to a generalization of the method presented by the author in~\cite{Chataig:2019}. We have not addressed the measurement problem or the origin of probabilities, but we argued that the relational content of a diffeomorphism-invariant quantum theory can be understood in terms of conditional probabilities. The formalism here presented is not meant to be the definitive method of construction and interpretation of observables in quantum gravity, but we believe it may prove useful in several toy models of quantum gravity (such as symmetry-reduced quantum cosmology) and, in particular, in the computation of quantum-gravitational effects in the early Universe. 

Relational observables describe the evolution of on-shell tensor fields with respect to each other in a diffeomorphism-invariant fashion. In the classical theory, such observables are constructed as diffeomorphism-invariant extensions of gauge-fixed components of tensor fields [i.e., the components written in a particular coordinate system defined by a gauge condition such as~(\ref{eq:gauge-condition-general})]. The interpretation of the classical relational observable $\Ob[f|\chi = s]$ is straightforward: it is the value of the field $f$ when the field $\chi$ is observed to have the value $s$. Thus, $\Ob[f|\chi = s]$ is a prediction conditioned on the value of $\chi$. But how do we construct these observables in the quantum theory, and in what sense are they relational?

A method of constructing the quantum version of relational observables was proposed by the author in~\cite{Chataig:2019} (see also~\cite{Hoehn:Trinity} for a similar approach). In the present article, we have presented a refined version of the method of~\cite{Chataig:2019}, which corresponds to building the quantum relational observables via their spectral decomposition [cf.~(\ref{eq:conditions-omega-1}) and~(\ref{eq:conditions-omega-2})]. As we have shown in Sec.~\ref{sec:gf-propagator}, once the eigenstates of relational observables are known, one can construct a gauge-fixed propagator, which dictates the unitary evolution of physical states with respect to the (in principle arbitrarily chosen) gauge-fixed time parameter $s$. The unitarity is a consequence of the fact that eigenstates of the relational observables form a complete orthonormal system in the physical Hilbert space for each value of $s$ if the gauge condition is well-defined, i.e., admissible according to the criterion discussed in Sec.~\ref{sec:relObs-I}. This completeness yields an operator version of the Faddeev-Popov resolution of the identity frequently used in path integrals. In analogy to the classical theory, the physical Hilbert space is divided into multiplicity $\sigma$ sectors, which are the generalization of the positive and negative frequency sectors of the quantum relativistic particle.

Moreover, we have discussed how this formalism can be applied to the case in which the gauge condition is canonically conjugate to an invariant self-adjoint Hamiltonian [cf.~Sec.~\ref{sec:relObs-II}]. In this case, an explicit formula for the Faddeev-Popov operator is available [cf.~(\ref{eq:FP-general-quantum-powers})], and we have derived the operator equations of motion for the relational observables, referred to as the gauge-fixed Heisenberg equations [cf.~(\ref{eq:gf-Heisenberg})]. We have shown that these equations hold for two choices of factor ordering in the definition of the quantum relational observables. The unitarity of the evolution associated with the gauge-fixed Heisenberg equations or with the gauge-fixed propagator shows that time and dynamics do not vanish in the quantum theory (``problem of time'') as is frequently claimed~\cite{Kiefer:book}. The measurement problem is, however, still present.

Nevertheless, the gauge-fixed time parameter $s$ (or, as described in~\cite{Hoehn:Trinity}, the reading of a clock)  is a c-number that requires a clear physical interpretation. This interpretation should also clarify the sense in which the quantum observables are really ``relational'' objects. This was not addressed in~\cite{Chataig:2019}, and the previous proposals~\cite{Rovelli:1990-1,Rovelli:1990-2,Rovelli:1991,Tambornino:2012} also do not seem to give a satisfactory interpretation. In this article, we have argued that, just as the classical relational observables can be seen as descriptions of the physical quantities (worldline tensors) in the time reference frame defined by the level sets of the gauge condition, the quantum relational observables define a notion of quantum reference frames [cf.~Sec.~\ref{sec:QRef}], in which the time parameter $s$ is defined from the spectrum of the gauge condition $\hat{\chi}$. More generally, we have argued that the overlap of physical states with the eigenstates of relational observables corresponds to the representation of the states in the reference frame defined by $\hat{\chi}$. In this way, a change of quantum reference frame corresponds to change of basis in the physical Hilbert space [cf.~Sec.~\ref{sec:switch}]. Thus, different reference frames can in principle be described in a single Hilbert space (this was also observed in the other approaches described in~\cite{Hoehn:Trinity,QRef3,QRef4,Barvinsky}). However, it is not sufficient to declare that $s$ is an eigenvalue of the gauge condition $\hat{\chi}$ or that $\hat{\chi}$ takes the definite classical value $s$ (in the appropriate reference frame) without any further explanation. One must relate $s$ to observations. Since the classical $\Ob[f|\chi = s]$ is a conditional prediction, it then seems reasonable to consider that its quantum version must be associated with conditional probabilities.

For this reason, we considered in Sec.~\ref{sec:CP-general} the definition of conditional probabilities from the solutions of the quantum constraint equation(s), and we have shown that the quantum averages of suitably defined relational observables are equivalent to conditional expectation values of the worldline tensors in definite multiplicity sectors (e.g., definite frequency sectors in the case of the relativistic particle). We believe this clarifies the physical interpretation of $s$ and, thus, the sense in which $\hat{\Ob}[f|\chi = s]$ is a relational object. Indeed, $s$ is the observed value of the field $\hat{\chi}$ (in an experiment), conditioned on which we can make probabilistic predictions about the values of $\hat{f}$. As shown in Sec.~\ref{sec:CP-general}, this information is equivalently encoded in the conditional probabilities or in the eigenstates of the quantum relational observables. Thus, $\hat{\Ob}[f|\chi = s]$ is a relational object in the sense that it allows one to make conditional predictions about the quantum fields in a generally covariant theory.

In this way, the quantum reference frame in which time is defined from the spectrum of $\hat{\chi}$ can also be defined from the space of conditional wave functions associated with the conditional probabilities [cf.~Sec.~\ref{sec:invext}]. We thus have two descriptions of the quantum dynamics in definite multiplicity sectors: the use of conditional probability amplitudes (referred to as the gauge-fixed point of view, since it represents the dynamics given that $\hat{\chi}$ is observed to have the value $s$) and the use of relational observables (referred to as the invariant point of view, as observables are diffeomorphism-invariant quantitites). The equivalence of these two points of view was first noted in~\cite{Hoehn:Trinity} and the present article can be seen an extension of~\cite{Hoehn:Trinity} adapted to the formalism of~\cite{Chataig:2019}. Moreover, we note that, after the submission of the present article, the authors of~\cite{Hoehn:Trinity} have released a generalization of their related formalism, which features similar conclusions and results that are technically complementary to the ones presented here~\cite{Hoehn:upcoming}.

We have also shown in Sec.~\ref{sec:PW} that the frequently used Page-Wootters formalism is a particular case of our approach and that our results can be seen as a generalization of the equivalence between the Page-Wootters method and the construction of relational observables which was found in~\cite{Hoehn:Trinity} for a particular class of models and gauge conditions. Furthermore, we have seen how our construction of quantum relational observables reduces in the Page-Wootters case to the $G$-twirl operation (related to the so-called relativization maps) that have been used in the quantum foundations literature~\cite{Hoehn:Trinity,QRef1,QRef3,QRef4,QRef5}. It would be interesting to apply our approach to different examples of time-reparametrization invariant quantum mechanics in an effort to generalize certain results already obtained with the Page-Wootters formalism. We hope to address this in the future.

In Sec.~\ref{sec:flrw}, we have analyzed the example of a recollapsing cosmology, for which we constructed the physical Hilbert space and the relevant quantum relational observables, which obey a unitary gauge-fixed Heisenberg equation of motion governed by an invariant Hamiltonian [cf.~(\ref{eq:flrw-gf-Heisenberg})]. We have also discussed how the relational quantum dynamics can be understood from the definition of conditional wave functions that are invariant extensions of relative initial data [cf.~Sec.~\ref{sec:flrw-rel-dynamics}]. This example illustrates how the formalism here described can be of use in quantum cosmology. Indeed, the present formalism is useful because it is directly applicable to solvable minisuperspace (symmetry-reduced) models of quantum cosmology and, more generally, to models of time-reparametrization invariant quantum mechanics. Such models are often analyzed as toy-models of quantum gravity or as attempts to describe quantum mechanics without an external time parameter.

While our focus has been on conceptual matters and on formalism, which is necessary if one is to obtain a consistent diffeomorphism-invariant quantum theory with a sensible interpretation, one must also face the question of extracting falsifiable predictions. What kind of predictions can be made? And are they relevant to cosmology? By assuming that all physical degrees of freedom should be described in a diffeomorphism invariant way, one is led to the view that physical observables are relational and, according to the formalism we have presented, that their quantum dynamics is encoded in conditional correlation functions. These are the quantities that can be predicted in the quantum theory. Moreover, the early Universe may be one of the epochs in which imprints of quantum gravity might have observable consequences. It is then pertinent to ask: (1) how does the formalism we present here relate to the usual observations and measurements in cosmology? (2) What is the connection between the (quantum) relational observables we have constructed and the usual cosmological observables? 

Our current cosmological measurements refer to a classical spacetime background, with respect to which the usual observables and primordial correlation functions are computed. Note that this is also a relational description: one may reinterpret the usual primordial correlation functions (and the usual cosmological observables) as relational quantities that are conditioned on the (``late-time'') classical values of the metric field. However, if one adopts a quantum description of spacetime in the early Universe, it is, in principle, possible to compute corrections to the dynamics of primordial correlators. These corrections would originate from the hypotheses: (1) there is no preferred, classical spacetime background in the early Universe; (2) the quantum dynamics is diffeomorphism invariant and relational. In this way, the formalism we present here would recover the description of the usual observations and measurements in cosmology in the ``late-time'' classical limit of the metric field, whereas all observables are relational. The primordial correlators and their corrections would be understood as conditional quantities. This program has been carried out in the recent article~\cite{ChataigKraemer}, in which a weak-coupling expansion is used to compute corrections of quantum-gravitational origin to the power spectra of primordial fluctuations in (quasi-)de Sitter space. The question of unitarity and observability of these corrections (e.g., in the cosmic microwave background spectrum) is also discussed in~\cite{ChataigKraemer}, and comments on further directions of research are given.

Finally, what is the relevance of these results for the construction and interpretation of diffeomorphism-invariant operators in quantum gravity? The classical diffeomorphism-invariant observables in general relativity are complicated, possibly nonlocal objects and their quantization is hardly a trivial matter. A generalization to field theory of the formalism here described could facilitate this issue. This is because of the equivalence between the gauge-fixed and invariant points of view explained above. Indeed, instead of working with the complicated relational observables (invariant point of view), one may choose to compute the often simpler conditional probabilities (gauge-fixed point of view) in a definite multiplicity sector. This equivalence would also provide the physical interpretation of the relational observables in field theory: as in the mechanical case, their eigenstates can be used to make conditional predictions.

As the formalism we have presented was restricted to mechanical models, its generalization to the field-theoretic setting would necessarily require a careful regularization of the quantum constraint equations. Moreover, one would need to ascertain whether the quantum constraint algebra is anomalous. These are nontrivial problems which are outside of the scope of this article. Nevertheless, if the quantum constraint algebra were to be successfully regularized and proven to be consistent, a generalization of the method here presented would be possible. In this case, the solutions to the quantum constraint equations would be used to define conditional probabilities associated with observations of tensor fields given certain gauge conditions. If one is content with computing conditional expectation values (i.e., if one is satisfied with working only in the gauge-fixed point of view), the construction of diffeomorphism-invariant relational observables would not be necessary. In this case, the quantum dynamics could be understood from the conditional predictions made directly from the on-shell wave functional, which is interpreted relationally as an invariant extension of relative initial data. We leave this fascinating topic for future work.

\begin{acknowledgments}
The author thanks Claus Kiefer for useful discussions that inspired this work, Philipp A. H\"{o}hn for comments on a first version of the manuscript, an anonymous referee for constructive criticism, and the Bonn-Cologne Graduate School of Physics and Astronomy for financial support.
\end{acknowledgments}


\begin{thebibliography}{99}

\bibitem{Oriti:book}
{\it Approaches to Quantum Gravity: Toward a New Understanding of Space, Time and Matter},
edited by D.~Oriti,
(Cambridge University Press, Cambridge, England, 2009).

\bibitem{PW1}
  D.~N.~Page and W.~K.~Wootters,
  Phys.\ Rev.\ D \href{https://doi.org/10.1103/PhysRevD.27.2885}{{\bf 27} 2885} (1983).

\bibitem{PW2}
W.~K.~Wootters,
Int. J. Theor. Phys. \href{https://doi.org/10.1007/BF02214098}{{\bf 23}, 701} (1984).

\bibitem{PW3}
  D.~N.~Page,
  CERN Report No. NSF-ITP-89-18, 1989.

\bibitem{PW4}
D.~N.~Page,
in {\it Physical Origins of Time Asymmetry},
edited by J. J. Halliwell, J. P\'{e}rez-Mercader, and W. H. Zurek,
(Cambridge University Press, Cambridge, England, 1994).

\bibitem{PW5}
 D.~N.~Page,
 in {\it Conceptual Problems of Quantum Gravity},
 edited by A.~Ashtekar and J.~Stachel,
 (Birkhauser, Boston, 1991).
 
\bibitem{PW6}
D.~N.~Page,
in {\it String Theory, Quantum Cosmology and Quantum Gravity: Proceedings of the Paris-Meudon Colloquium},
edited by H~DeVega and N.~Sanchez,
(World Scientific, Singapore, 1987).

\bibitem{PW7}
D.~N.~Page,
in {\it Gravitation: A Banff Summer Institute},
edited by R.~Mann and P.~Wesson,
(World Scientific, Singapore, 1991).

\bibitem{Dolby:2004}
  C.~E.~Dolby,
  \href{https://arxiv.org/abs/gr-qc/0406034}{arXiv:gr-qc/0406034}.

\bibitem{Hoehn:Trinity}
  P.~A.~H\"{o}hn, A.~R.~H.~Smith and M.~P.~E.~Lock,
  \href{https://arxiv.org/abs/1912.00033}{arXiv:1912.00033 [quant-ph]}.

\bibitem{DH2}
  R.~B.~Griffiths,
  {\it Consistent Quantum Theory},
  (Cambridge University Press, Cambridge, England, 2008).

\bibitem{DH4}
R.~Omn\`{e}s, {\it The Interpretation of Quantum Mechanics}, (Princeton University Press, Princeton, NJ, 1994).

\bibitem{DH5}
M.~Gell-Mann and J.~B.~Hartle,
in {\it Complexity, Entropy, and the Physics of Information},
edited by W. Zurek,
SFI Studies in the Sciences of Complexity Vol. VII,
(Addison-Wesley, Reading, MA, 1990);
and in {\it Proceedings of the 3rd International Symposium on the Foundations of Quantum Mechanics in the Light of New Technology},
edited by S.~Kobayashi, H.~Ezawa, M.~Murayama, and S.~Nomura, (Physical Society of Japan, Tokyo, 1990).

\bibitem{DH6}
J.~B.~Hartle,
in {\it Gravitation and Quantizations, Proceedings of the 1992 Les Houches Summer School},
edited by B.~Julia and J.~Zinn-Justin,
(North Holland, Amsterdam, 1995).

\bibitem{DH8}
  D.~A.~Craig and P.~Singh,
  Phys.\ Rev.\ D \href{https://doi.org/10.1103/PhysRevD.82.123526}{{\bf 82} 123526} (2010).

\bibitem{relQM}
  C.~Rovelli,
  Int.\ J.\ Theor.\ Phys.\  \href{https://doi.org/10.1007/BF02302261}{{\bf 35} 1637} (1996).

\bibitem{dBB1}
  N.~Pinto-Neto and J.~C.~Fabris,
  Classical Quantum Gravity\  \href{https://doi.org/10.1088/0264-9381/30/14/143001}{{\bf 30} 143001} (2013).

\bibitem{dBB2}
  N.~Pinto-Neto and W.~Struyve,
  \href{https://arxiv.org/abs/1801.03353}{arXiv:1801.03353}.

\bibitem{Rovelli:1990-1}
  C.~Rovelli,
  Phys.\ Rev.\ D \href{https://doi.org/10.1103/PhysRevD.42.2638}{{\bf 42} 2638} (1990).

\bibitem{Rovelli:1990-2}
  C.~Rovelli,
  Classical Quantum Gravity\  \href{https://doi.org/10.1088/0264-9381/8/2/011}{{\bf 8} 297} (1991);
    \href{https://doi.org/10.1088/0264-9381/8/2/012}{{\bf 8} 317} (1991).

\bibitem{Rovelli:1991}
  C.~Rovelli,
  Phys.\ Rev.\ D \href{https://doi.org/10.1103/PhysRevD.43.442}{{\bf 43}, 442} (1991).
  
  \bibitem{Woodard:1985}
  N.~C.~Tsamis and R.~P.~Woodard,
  Classical Quantum Gravity\  \href{https://doi.org/10.1088/0264-9381/2/6/011}{{\bf 2} 841} (1985).

\bibitem{Teitelboim:1992}
  C.~Teitelboim,
  in {\it Physical Origins of Time Asymmetry},
edited by J. J. Halliwell, J. P\'{e}rez-Mercader, and W. H. Zurek,
(Cambridge University Press, Cambridge, England, 1994).

\bibitem{HT:book}
M.~Henneaux and C.~Teitelboim,
{\it Quantization of Gauge Systems},
(Princeton University Press, Princeton, NJ, 1992).

\bibitem{Woodard:1993}
  R.~P.~Woodard,
  Classical Quantum Gravity\  \href{https://doi.org/10.1088/0264-9381/10/3/008}{{\bf 10} 483} (1993).

\bibitem{Dittrich:2004}
  B.~Dittrich,
  Gen.\ Relativ.\ Gravit.\  \href{https://doi.org/10.1007/s10714-007-0495-2}{{\bf 39} 1891} (2007).

\bibitem{Dittrich:2005}
  B.~Dittrich,
  Classical Quantum Gravity\  \href{https://doi.org/10.1088/0264-9381/23/22/006}{{\bf 23} 6155} (2006).

\bibitem{Chataig:2019}
L.~Chataignier,
Phys. Rev. D \href{https://doi.org/10.1103/PhysRevD.101.086001}{\textbf{101} 086001} (2020).

\bibitem{Tambornino:2012}
  J.~Tambornino,
  SIGMA \href{https://doi.org/10.3842/SIGMA.2012.017}{{\bf 8} 017} (2012).

\bibitem{Hoehn:2018-1}
  A.~Vanrietvelde, P.~A.~H\"{o}hn, F.~Giacomini, and E.~Castro-Ruiz,
  Quantum \href{https://doi.org/10.22331/q-2020-01-27-225}{{\bf4}, 225} (2020).

\bibitem{Hoehn:2018-2}
  P.~A.~H\"{o}hn and A.~Vanrietvelde,
  \href{https://arxiv.org/abs/1810.04153}{arXiv:1810.04153}.

\bibitem{Hoehn:2019}
  P.~A.~H\"{o}hn,
  Universe \href{https://doi.org/10.3390/universe5050116}{{\bf 5},  116} (2019).

\bibitem{Kiefer:book}
C.~Kiefer,
{\it Quantum Gravity}, 3rd ed.,
International Series of Monographs on Physics 
(Oxford University Press, Oxford, 2012).

\bibitem{QRef1}
S.~D.~Bartlett, T.~Rudolph, and R.~W.~Spekkens,
Rev. Mod. Phys. \href{https://doi.org/10.1103/RevModPhys.79.555}{\textbf{79} 555} (2007).

\bibitem{QRef3}
T.~Miyadera, L.~Loveridge, and P.~Busch,
J. Phys. \href{https://doi.org/10.1088/1751-8113/49/18/185301}{A {\bf49} 185301} (2016).

\bibitem{QRef4}
L.~Loveridge, T.~Miyadera, and P.~Busch,
Found. Phys. \href{https://doi.org/10.1007/s10701-018-0138-3}{{\bf48} 135} (2018).

\bibitem{QRef5}
P.~Busch, P.~Lahti, J.P.~Pellonp\"{a}\"{a}, and K.~Ylinen,
{\it Quantum Measurement},
Theoretical and Mathematical Physics, (Springer International Publishing, Switzerland, 2016).
 
\bibitem{HT:Vergara}
M.~Henneaux, C.~Teitelboim and J.~Vergara,
Nucl. Phys. \href{https://doi.org/10.1016/0550-3213(92)90166-9}{\textbf{B387} 391} (1992).
 
\bibitem{Marolf:1995}
  D.~Marolf,
  Classical Quantum Gravity\  \href{https://doi.org/10.1088/0264-9381/12/5/011}{{\bf 12} 1199} (1995).

\bibitem{FP-1}
  L.~D.~Faddeev and V.~N.~Popov,
  Phys.\ Lett.\  \href{https://doi.org/10.1016/0370-2693(67)90067-6}{{\bf 25B} 29} (1967).

\bibitem{FP-2} 
  L.~D.~Faddeev and V.~N.~Popov,
  Sov.\ Phys.\ Usp.\  \href{https://doi.org/10.1070/PU1974v016n06ABEH004089}{{\bf 16}, 777} (1974)
  [Usp.\ Fiz.\ Nauk {\bf 111}, 427 (1973)].

\bibitem{Rieffel:1974}
M.~A.~Rieffel,
Adv. Math. \href{https://doi.org/10.1016/0001-8708(74)90068-1}{{\bf 13}, 176} (1974).

\bibitem{HT-SUSY:1982}
  M.~Henneaux and C.~Teitelboim,
  Ann. Phys. (N.Y.)\  \href{https://doi.org/10.1016/0003-4916(82)90216-0}{{\bf 143} 127} (1982).

\bibitem{Landsman:1995}
N.~P.~Landsman,
J. Geom. Phys. \href{https://doi.org/10.1016/0393-0440(94)00034-2}{{\bf15}, 285} (1995).

\bibitem{Marolf:1995-4}
  D.~Marolf,
  \href{https://arxiv.org/abs/gr-qc/9508015}{arXiv:gr-qc/9508015}.
  
\bibitem{Marolf:1997}
  J.~B.~Hartle and D.~Marolf,
  Phys.\ Rev.\ D \href{https://doi.org/10.1103/PhysRevD.56.6247}{{\bf 56}, 6247} (1997).
  
\bibitem{Marolf:2000}
  D.~Marolf,
  \href{https://arxiv.org/abs/gr-qc/0011112}{arXiv:gr-qc/0011112}.

\bibitem{Pauli}
W.~Pauli,
{\it General Principles of Quantum Mechanics},
translated by P.~Achuthan and K.~Venkatesan
(Springer, Berlin 1980).

\bibitem{Barvinsky}
A.~Barvinsky,
Phys. Rep. \href{https://doi.org/10.1016/0370-1573(93)90032-9}{\textbf{230} 237} (1993).

\bibitem{ChataigKraemer}
L.~Chataignier and M.~Kr\"{a}mer,
\href{https://arxiv.org/abs/2011.06426}{arXiv:2011.06426}.

\bibitem{Hunter:1975}
  G.~Hunter,
   Int. J. Quantum Chem. \href{https://doi.org/10.1002/qua.560090205}{{\bf 9}, 237} (1975).
   
\bibitem{Kiefer:1993}
  C.~Kiefer,
  Lect.\ Notes Phys.\  \href{https://doi.org/10.1007/3-540-58339-4_19}{{\bf 434} 170} (1994).
   
\bibitem{Chataig:2019-0}
    L.~Chataignier,
    Z.\ Naturforsch.\ A \href{https://doi.org/10.1515/zna-2019-0223}{{\bf 74} 1069} (2019).

\bibitem{Gambini:2009}
  R.~Gambini, R.~A.~Porto, J.~Pullin, and S.~Torterolo,
  Phys.\ Rev.\ D \href{https://doi.org/10.1103/PhysRevD.79.041501}{{\bf 79} 041501} (2009).

\bibitem{Kiefer:1988}
  C.~Kiefer,
  Phys.\ Rev.\ D \href{https://doi.org/10.1103/PhysRevD.38.1761}{{\bf 38} 1761} (1988).

\bibitem{Palais:1979} 
  R.~S.~Palais,
  Commun.\ Math.\ Phys.\  \href{https://doi.org/10.1007/BF01941322}{{\bf 69} 19} (1979).

\bibitem{Torre:1999}
  C.~G.~Torre,
  Int.\ J.\ Theor.\ Phys.\  \href{https://doi.org/10.1023/A:1026650212053}{{\bf 38} 1081} (1999)

\bibitem{Fels:2002}
  M.~E.~Fels and C.~G.~Torre,
  Classical Quantum Gravity\  \href{https://doi.org/10.1088/0264-9381/19/4/303}{{\bf 19} 641} (2002).

\bibitem{Bessel}
T.~M.~Dunster, SIAM J. Math. Anal. \href{https://doi.org/10.1137/0521055}{{\bf 21} 995} (1990);
A.~Passian, H.~Simpson, S.~Kouchekian, and S.~B.~Yakubovich,
J. Math. Anal. Appl. \href{https://doi.org/10.1016/j.jmaa.2009.06.067}{{\bf 360} 380} (2009);
R.~Szmytkowski and S.~Bielski,
J. Math. Anal. Appl. \href{https://doi.org/10.1016/j.jmaa.2009.10.035}{{\bf 365} 195} (2010).

\bibitem{DeWitt:1967}
  B.~S.~DeWitt,
  Phys.\ Rev.\  \href{https://doi.org/10.1103/PhysRev.160.1113}{{\bf 160} 1113} (1967).
  
\bibitem{Hoehn:upcoming}
P.~A.~H\"{o}hn, A.~R.~H.~Smith, and M.~P.~E.~Lock,
\href{https://arxiv.org/abs/2007.00580}{arXiv:2007.00580}.

\end{thebibliography}
\end{document}